\documentclass{aa} 

\usepackage{mathabx}
\usepackage{txfonts}
\usepackage{graphicx}	
\usepackage{amsmath}	
\usepackage{amssymb}	
\usepackage{multirow}   
\usepackage{array}      %
\usepackage{mathtools}
\usepackage{xcolor} 
\usepackage[none]{hyphenat}

\usepackage[pdfpagelabels=false]{hyperref}

\hypersetup{colorlinks=true,linkcolor=blue,citecolor=blue,filecolor=blue,urlcolor=blue,}
\makeatletter
\renewcommand*\aa@pageof{, page \thepage{} of \pageref*{LastPage}}
\makeatother

\definecolor{green_comm}{RGB}{0,160,0}
\definecolor{orange}{RGB}{255,165,0}
\graphicspath{./Figures/}

\newcommand{\de}{\partial}
\newcommand{\der}[2]{\displaystyle \frac{\de #1}{\de #2}}
\newcommand{\Hy}{\ion{H}{}}
\newcommand{\HI}{\ion{H}{i}}
\newcommand{\HII}{\ion{H}{ii}}
\newcommand{\He}{\ion{He}{}}
\newcommand{\HeI}{\ion{He}{i}}
\newcommand{\HeII}{\ion{He}{ii}}
\newcommand{\HeIII}{\ion{He}{iii}}
\newcommand{\HeITR}{\ion{He}{i}(2\,$^3$S)}

\newcommand{\FWHM}{\text{FWHM}}
\newcommand{\mhe}{\ion{He}{i}(2\,$^3$S)}
\begin{document}

\title{Self-Consistent Modeling of Metastable Helium Exoplanet Transits}
   \titlerunning{\HeITR~ transits}
   \authorrunning{Biassoni et al.}

\author{Federico Biassoni\inst{1,2} 
        \and Andrea Caldiroli \inst{3} 
        \and Elena Gallo\inst{4}
        \and Francesco Haardt\inst{1,2,5}
        \and Riccardo Spinelli\inst{6}
        \and Francesco Borsa \inst{2}
}

\institute{
        DISAT, Universit\`a degli Studi dell'Insubria, via Valleggio 11, I-22100 Como, Italy
\and
        INAF -- Osservatorio Astronomico di Brera, Via E. Bianchi 46, 23807 Merate, Italy
\and            
        Fakult\"at f\"ur Mathematik, Universit\"at Wien, Oskar-Morgenstern-Platz 1, A-1090 Wien, Austria
\and
        Department of Astronomy, University of Michigan, 1085 S University, Ann Arbor, Michigan 48109, USA
\and
        INFN, Sezione Milano-Bicocca,P.za della Scienza 3, I-20126 Milano, Italy
\and
        INAF – Osservatorio Astronomico di Palermo, Piazza del Parlamento 1, I-90134, Palermo, Italy
             }

\date{Received; accepted}

\abstract
{Absorption of stellar X-ray and Extreme Ultraviolet (EUV) radiation in the upper atmosphere of close-in exoplanets can give rise to hydrodynamic outflows, which may lead to the gradual shedding of their primordial, light element envelopes. Excess absorption by neutral helium atoms in the metastable 2\,$^3$S state [\mhe], at $\sim$10,830 \r{A}, has recently emerged as a viable diagnostic of atmospheric escape. Here we present a public add-on module to the 1D photo-ionization hydrodynamic code ATES, designed to calculate the \mhe\ transmission probability for a broad range of planetary parameters. By relaxing the isothermal outflow assumption, the code enables a self-consistent assessment of the \mhe\ absorption depth along with the atmospheric mass loss rate and the outflow temperature {profile}, which strongly affects the recombination rate of \HeII~into \mhe.
We investigate how the transit signal can be expected to depend upon {known} system parameters, including host spectral type, orbital distance, as well as planet gravity. At variance with previous studies, which identified K-type stars as favorable hosts, we conclude that late M-dwarfs with Neptune-sized planets orbiting at $\sim$0.05-0.1 AU can be expected to yield the strongest transit signal -- well in excess of 30\% for near-cosmological He-to-H abundances. More generally, we show that the physics which regulates the population and depletion of the metastable state, combined with geometrical effects, can yield somewhat counter-intuitive results, such as a non-monotonic dependence of the transit depth on orbital distance. These are compounded by a strong degeneracy between the stellar EUV flux intensity and the atmospheric He-to-H abundance, both of which are highly uncertain. Compared against spectroscopy data, now available for over 40 systems, our modelling suggests that either a large fraction of the targets have helium depleted envelopes, or, that the input stellar EUV spectra are systematically overestimated. The updated code and transmission probability module are available publicly as an online repository.
}

\titlerunning{Metastable Helium Transits}
\maketitle
%

\section{Introduction}
\label{sec:intro}
Absorption of high-energy stellar radiation in the upper atmosphere of close-in exoplanets can lead to the formation of hydrodynamic outflows, and the removal of a substantial portion of the atmosphere's primordial, light element envelope \citep{Vidal_2003,Yelle2004,Tian2005,Koskinen2014,Kubyshkina_2018}.
Hydrodynamic escape is thought to play a significant role in shaping the observed properties of the Kepler planet population, likely contributing to carving the observed radius valley \citep{fulton17} by stripping planets of their initial envelopes, and turning them into rocky cores \citep{owenwu13,owenwu17}. 

Transit spectroscopy of light elements indeed provides evidence for evaporating atmospheres; hydrogen Ly$\alpha$ absorption signatures have been detected in a handful of transiting hot Jupiters and Neptunes (HD 209458b \citealt{Vidal_2003, Ben_Jaffel_2007}; HD 189733b \citealt{Lecavelier_2010}; GJ 436b \citealt{Kulow_2014, Ehrenreich_2015} and GJ 3470b \citealt{Bourrier_2018}). 
The reader is referred to \cite{Owen_2023} for a comprehensive review. 
Reconstructing the atmospheric Ly$\alpha$ feature, however, is extremely challenging. Combined with the Earth's geocoronal Ly$\alpha$ emission, interstellar scattering makes the core of the line (in the velocity range $\pm 30$ km sec$^{-1}$) practically inaccessible. ``Line'' detections are effectively limited to its Doppler wings, whose high (typically blue-shifted) velocities likely probe regions far from the outflow thermal launching site, where the escaping material is bound to be accelerated by the stellar wind \citep{Vidal_2003}. \\

\cite{Seager_2000} first identified \mhe\ as an additional, promising absorption feature for exoplanet transmission spectroscopy. The feature arises from recombination of \HeII\ to the metastable\footnote{Being radiatively decoupled from the ground (singlet) state, \mhe\ is metastable, with a lifetime of $\simeq$ 2.18 hr \citep{Drake_1971}. } 2\,$^3$S triplet state, and subsequent radiative transitions to the 2\,$^3$P state (at 10,830.34, 10,830.25, and 10,829.09 \r{A}, albeit the doublet at $\sim$10,830 \r{A} dominates the signal). 
More recently, \citet{Oklopcic_2018} developed a 1D isothermal model for escaping hydrogen-helium atmospheres, where the equations for the neutral helium fraction are written separately for the singlet and triplet state. They apply this model to the specific case of GJ 436b and HD 209458b (both with detected, escaping Ly$\alpha$), demonstrating the detectability of excess \mhe\ absorption during transits (at the 8 and 2\% level, respectively). \mhe\ absorption is also relevant in the context of a recent development; the possibility to use this feature to infer planetary magnetic fields, either with spectropolarimetry \citep{Oklopcic2020}, or via line shifts \citep{Schreyer_2023}.\\

Unlike Ly$\alpha$, \mhe\ does not suffer from massive interstellar absorption (as the metastable state is populated through helium ionization and subsequent recombination), and it is also observable from the ground. 
A dozen detections have been reported so far; WASP-107b \citep{Spake_2018}; HAT-P-11b \citep{Mansfield_2018,Allart_2018}; WASP-69b \citep{Nortmann_2018}; HD 189733b \citep{Salz_2018, Guilluy_2020}; HD 209458b \citep{Alonso_2019}; GJ 3470 b \citep{Palle2020,Ninan2020}; HAT-P-18 b \citep{Paragas2021}; TOI-560 b \citep{Zhang2022}. 
The reader is referred to \cite{Bennett_2023}, and references therein, for a recent summary of the observational work pertaining \mhe\ transit spectroscopy, including an ever-growing list of non-detections, for WASP-48b, GJ 436b \citep{Nortmann_2018}; WASP-80b \citep{fossati22}; HD 97658b \citep{Kasper2020}; GJ 9827d \citep{Kasper2020}; GJ 1214b \citep{Kasper2020,Orell-Miquel2022}; WASP-12 b \citep{Kreidberg2018}; WASP-127 b \citep{dosSantos2020}; KELT-9 b \citep{Sanchez2022}; 55 Cnc e \citep{Zhang2021} and the  V1298 Tau system \citep{vissa21}. To this, \citep{Guilluy_2023} recently added spectroscopic data for HAT-P-3b, HAT-P-33b, HAT-P-49b, HD89345b, K2-105b, Kepler-25c, Kepler-63b, Kepler-68b, and WASP-47d, none of which are detected. \\

As for Ly$\alpha$, non detections --particularly in young/active and/or highly irradiated planets-- are especially important to understand.   
\citet{Poppenhaeger_2022} pointed out that most of the narrow band EUV stellar flux that ionizes helium (the first step toward populating the metastable state) arises from individual emission lines, mostly by iron at coronal temperatures, and showed that the large scatter in the helium transit depth to stellar activity relation \citep{Bennett_2023} can be reduced by taking coronal iron abundance into account. Nevertheless, the non-detections of extreme system, such as WASP-80b, remains puzzling.

Since hydrodynamic escape is driven primarily by EUV photons, the lack of observational data in this range severely hampers our ability to make accurate predictions for the atmospheric mass outflow rates and ionization properties.
Broadband EUV observations that are virtually unaffected by absorption are only available for the Sun. In addition, EUV data are only available for a dozen stars of type F and later, from the Extreme UltraViolet Explorer (EUVE; \citealt{craig97}). Lacking system-specific data, stellar EUV spectra of planet hosts are either modeled by using the EUVE spectra of similar type stars, or, approximated on the basis of scaling relations with Ly$\alpha$ or X-rays \citep{Chadney_2005,Linsky_2014,King_2018}.
This is bound to yield substantial uncertainties.

An additional source of uncertainty is the helium abundance in the upper atmosphere. As an example, the 0.7\% upper limit (at 2$\sigma$) for the hot Jupiter WASP-80b can be interpreted as due to a highly sub-solar He abundance \citep{fossati22}, or by low EUV irradiation by the host star \citep{Fossati_2023}. When modeling observational data, the He abundance vs. EUV flux degeneracy is often compounded by the choice of isothermal outflow models (or Parker winds; \citealt{Parker_1958}), where the atmospheric mass outflow rate and temperature are considered as model parameters that can be tweaked to reproduce the observed transit spectroscopy feature (or lack thereof). An additional complication arises from the fact that hydrodynamic escape itself may enhance the atmospheric helium-to-hydrogen fraction \citep{Malsky_2023}. \\

The primary goal of this work is to develop an efficient and reliable model that enables a systematic study of metastable helium transits over a broad parameter range, and where the hydrodynamic outflow solution, i.e., mass outflow rate and temperature, is calculated self-consistently with the \mhe\ density profile and expected transit absorption excess. 

This is accomplished by modifying the public 1D photoionization hydrodynamic code ATES \citep{Caldiroli_2021} to solve the continuity equations separately for the neutral helium triplet and singlet (Sect. \ref{sec:chem}), and by developing a dedicated transmission spectroscopy module to calculate the expected line absorption depth during transits (Sect. \ref{sec:TPM}).  We use the new code and module to assess the impact of stellar host spectral type, planet gravity and orbital distance on the expected \mhe\ transit signal (Sect. \ref{sec:results}).  By relaxing the isothermal outflow assumption, we can self-consistent model the outflow properties concurrently with the atmospheric absorption signal.\\

Throughout, we refer to photons in the range between 4.8-13.6 eV (912-2,585 \r{A}) as Far Ultraviolet (FUV); photons between 13.6-124 eV (100-912 \r{A}) are referred to as EUV radiation, whereas XUV indicates EUV plus X-ray photons, in excess of 124 eV (10-912 \r{A}). We are bound to consider only planets with optical radii in excess of 1.5 $R_{\Earth}$ (or 0.134 $R_J$), which are likely to retain primordial (i.e., hydrogen-helium rich) atmospheres \citep{Lopez2014}.

\begin{figure}
    \centering
    \includegraphics[width =0.9\columnwidth]{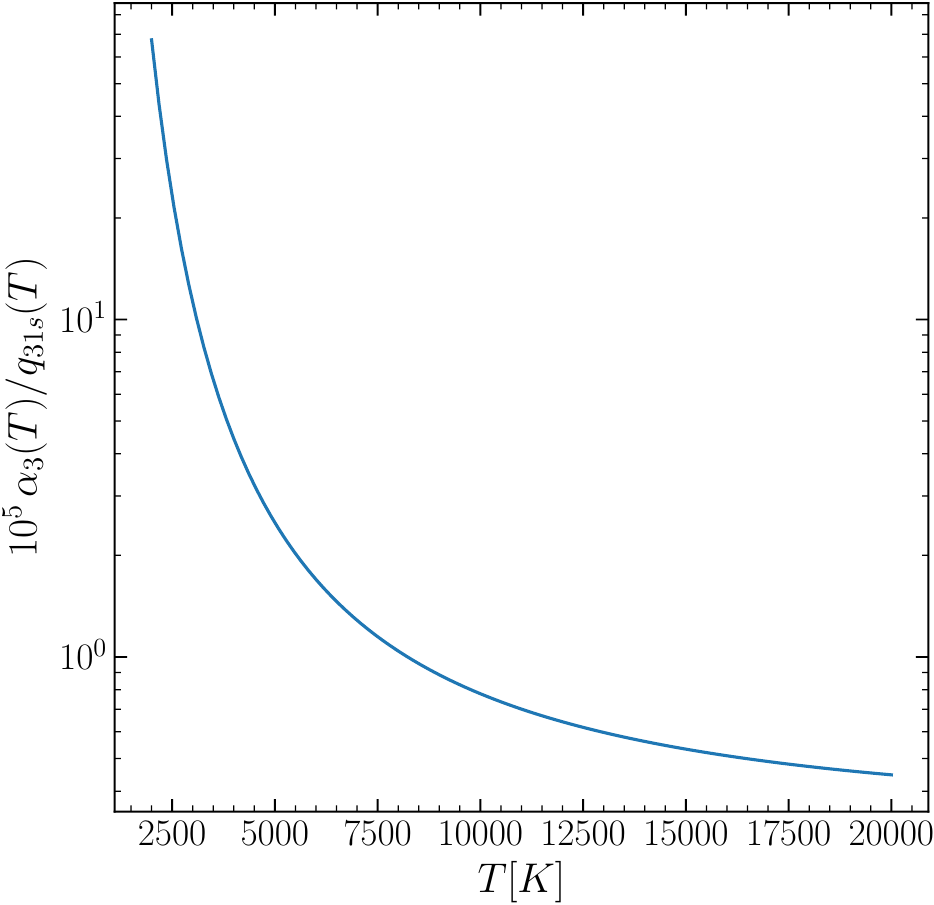}
    \caption{Ratio of the \HeII~to \HeITR~recombination coefficient [$\alpha_3(T)$] to the \HeITR~to HeI($2^1$S) collisional excitation rate [$q_{31s}(T)$], as a function of the flow temperature.}
    \label{fig:rec_ratio}
\end{figure}

\section{Methodology}
\label{sec:ATES_updates}
Our starting point is the 1D photoionization hydrodynamics code ATES \citep{Caldiroli_2021}. Given the planetary system's parameters (stellar irradiation spectrum\footnote{ATES can either assume a power-law-like stellar spectrum, or read in a binned numerical spectrum provided by the user.}, planet mass, radius, and equilibrium temperature, orbital distance, equilibrium temperature, stellar mass and envelope H/He abundance), ATES computes the temperature, density, velocity, and ionization fraction profiles of the irradiated planetary atmosphere, along with the instantaneous, steady-state mass loss rate. 
The outflow extends from the planet optical radius, $R_P$, to the system's Hill radius, $R_H$\footnote{The effective Roche lobe radius ($R_R\simeq 2/3 R_H$) is occasionally used in the literature in place of the Hill radius \citep[see, e.g.,][]{Guilluy_2023}. Using $R_R$, and assuming a constant \HeITR~density profile, yield a $\approx 50\%$ lower absorption depth.}.
All the simulations discussed below are initialized with a total density of $10^{14}$ g cm$^{-3}$ at the inner boundary; as discussed in section 3.6 of \cite{Salz2016a} this choice introduces a $<$ 50\% error in the mass outflow rates. The outer boundary choice, on the other hand, has non-negligible effects on the expected \mhe\ transit absorption, particularly in comparison to isothermal models; we come back to this point in Sect. ~\ref{sec:conclusions}.

The code solves the one-dimensional Euler, mass, and energy conservation equations in radial coordinates through a finite-volume scheme. The photoionization equilibrium solver includes cooling via bremsstrahlung, recombination, plus collisional excitation and ionization. Ion advection is implemented in post-processing. The (user-specified) \He/\Hy\ ratio is kept constant throughout a given simulation run. With the exception of Sect. \ref{subsec:He}, all the simulations presented in this paper adopt a cosmological number density ratio: He/H=0.083.\\

First, we extend the main code chemical network to include both the singlet and triplet states of \HeI, although cooling and heating terms due to triplet transitions are neglected, since the density of the helium atoms in this state is 5 to 7 orders of magnitude lower than for the singlet. Second, we develop a transmission probability module (TPM), i.e., an add-on routine that makes use of ATES' solutions to compute the expected \HeITR~absorption signal during planetary transits. These are described in turn below. 

\subsection{Metastable Helium Triplet Chemistry}
\label{sec:chem}
 
Following \citet{Oklopcic_2018}, the stationary continuity equations for the neutral helium fraction are written separately for the singlet ($f_{\HeI_1}\equiv n_{\HeI_1}/n_\He$) and triplet ($f_{\HeI_3}\equiv n_{\HeI_3}/n_\He$) states:
\begin{equation}
    \begin{cases}
        \begin{aligned}
            v\der{f_{\HeI_1}}{r} = & -\Gamma_{\HeI_1} f_{\HeI_1} + f_{\HeI_3}(A_{31} + n_\HI Q_{31}) \\
                                   & + \left[f_\HeII \alpha_1 - f_{\HeI_1} q_{13s} + (q_{31s} + q_{31p})f_{\HeI_3}\right]n_e, \\ 
            v\der{f_{\HeI_3}}{r} = & -\Gamma_{\HeI_3} f_{\HeI_3} - f_{\HeI_3}(A_{31} + n_\HI Q_{31}) \\
                                   & + \left[f_\HeII \alpha_3 + f_{\HeI_1} q_{13s} - (q_{31s} + q_{31p})f_{\HeI_3}\right]n_e, 
        \end{aligned}
    \end{cases}
    \label{eq:HeTR_model_eq}
\end{equation}
where $v$ is the flow velocity, $\Gamma$ and $\alpha$ are the photo-ionization and recombination rate coefficients, $A_{31}$ is the radiative transition rate between the metastable and the ground state; $q_{ij}$ are the transition rates between level $i$ and $j$ (with $s$ referring to transition to S-states and $p$ to P-states) after collisions with free electrons of number density $n_e$, and $Q_{ij}$ is the transition rate after collisions with neutral hydrogen atoms of number density $n_\HI$.
Eqs. \ref{eq:HeTR_model_eq} are complemented by charge conservation; at variance with \cite{Oklopcic_2018}, we also include contributions to $n_e$ from \HeIII: $n_e = n_\HII + n_\HeII + 2n_\HeIII$.

For the temperature-dependent reaction coefficients, we adopt the same notation and functional forms of \citet{Lampon2020}; for the photoionization rates we adopt the formalism of \citet{Caldiroli_2021} (see their eq. 6). 
We approximate the photoionization cross-section of \mhe\ by fitting a piecewise power-law to the experimental data of \citet{Norcross1971};
the best-fitting coefficients are reported in Table~\ref{tab:HeITR_PhotoionCS_fit}.

\begin{table} 
    \centering
    \renewcommand{\arraystretch}{1.5}
    \caption{Best-fitting coefficients for the \HeI($2^3$S) photoionization cross section.}    \label{tab:HeITR_PhotoionCS_fit}
    \begin{tabular}{c|c|c}
         Energy interval [eV] & $\log_{10}(\sigma_{0} [{\rm cm}^2])$  & k \\
         \hline
         %
         4.8  - 7.5  &  0.6859  & -0.8134  \\
         7.5  - 35.0 &  0.8710  & -1.7720  \\
         35.0 - 46.0 & -8.4719  &  9.1038  \\
         46.0 - 60.0 &  3.3997  & -3.0390  \\
         \hline
    \end{tabular}
    \tablefoot{A sequence of cojoined power-laws of the form $\sigma_{\HeI_3}(E) = \sigma_{0}(E/4.8{\rm eV})^k$ is assumed over different energy ranges above the helium ionization energy of 4.8 eV; the cross section has units of $10^{-18}$~cm$^2$. 
    }
\end{table}
At each time step, the chemistry network determines the fractions of neutral helium in the singlet and triplet state by solving numerically the static limit (i.e. $v=0$) of Equations \ref{eq:HeTR_model_eq}. The advection terms are accounted for in post-processing (see Sect. 2.3 in \citealt{Caldiroli_2021}). 

We wish to emphasize that, under typical circumstances, the dominant mechanisms that populate and deplete the \mhe\ state are, respectively, recombinations from \HeII\ (set by $\alpha_3$) and electron collisions (set by $q_{31s}$). Assuming (i) ionization equilibrium, and (ii) $f_\HeII\simeq 1$, the neutral helium fraction in the triplet state can be approximated as the ratio between the recombination coefficient and collisional excitation rate: 
$f_{\HeI_3} \simeq \alpha_3/q_{31s}$. 
Figure~\ref{fig:rec_ratio} illustrates the temperature dependence of $f_{\HeI_3}$. In the 2,000-10,000 K range that is relevant to atmospheric outflows, we can expect the \HeITR~absorption feature to be a steeply decreasing function of the outflow temperature. As a consequence, any deviation from isothermal profiles will strongly affect the expected \mhe\ density, and the ensuing absorption signal. 

\subsection{Transmission Probability Module}
\label{sec:TPM}

Using as input the radial outflow properties calculated by ATES, 
the Transmission Probability Module (TPM) computes the expected \mhe\ transit signal; in addition to the parameters needed for the main code, the TPM must be provided with the extent of the stellar radius, $R_{\star}$. 

Assuming spherical symmetry, the transmission probability at wavelength $\lambda$, averaged over the extent of the stellar disk on the plane of the sky, can be written as:
\begin{equation}
   \langle T \rangle _\lambda = \frac{1}{\pi(R_{\rm max}^2 - R_P^2)} \int_{R_P}^{R_{\rm max}}{db\, 2\pi b\,\, {\rm e}^{-\tau_\lambda(b)}}. 
   \label{eq:trans}
\end{equation}
where the integral extends from the planet optical radius to $R_{\rm max}\equiv \min(R_H,R_\star)$; 
$\tau_\lambda(b)$ is the \mhe\ optical depth along a line of sight (l.o.s.) with impact parameter $b$, measured from the planet center:
\begin{equation}
    \tau_\lambda(b) = \int_{0}^{2\sqrt{R_H^2-b^2}}{dx\,n_{\HeI_1}(x)\, 
    \sigma_\lambda(x)}, 
    \label{eq:tau}
\end{equation}
%
For each triplet line, the (temperature and bulk velocity dependent) absorption cross-section $\sigma$ can be written as:
\begin{equation}
    \sigma_\nu = \frac{\sqrt{\pi} e^2 f_i}{4 \pi \epsilon_0 m_e c \Delta{\nu_i}}H(a_i,\chi_i),
    \label{eq:VoigtCrossSection}
\end{equation}
where $\nu_i$ is the line center frequency, and $f_i$ is the oscillator strength given by the {\fontfamily{pcr}\selectfont NIST}\footnote{\url{https://www.nist.gov/pml/atomic-spectra-database}} database, and 
$\Delta{\nu_i} = \nu_i\,{\rm v}_{\rm th}/c$, where the thermal velocity is ${\rm v}_{\rm th} = \sqrt{2k_BT/m_{\rm He}}$. $H(a_i,\chi_i)$ is the Voigt function, with $a_i = A_i/(4\pi\Delta\nu_i)$ and $\chi_i = \left(\nu - \nu_i\right)/\Delta\nu_i$. 
$A_i$ is the Einstein coefficient of the transition, taken again from {\fontfamily{pcr}\selectfont NIST}. The Voigt function is computed as the real part of the Faddeeva function of argument $z_i = \chi_i - {\rm v}_{\rm los}/{\rm v}_{\rm th} + \sqrt{-1}\,a_i$, where ${\rm v}_{\rm los}$ is the l.o.s. component of the outflow velocity. 

Last, the atmospheric transmission probability is normalized to the fraction of the stellar disk that is not covered by the atmosphere, whilst also accounting for the (100\% opaque) planetary disk within $R_P$: 

\begin{align}
    T_\lambda = \frac{R_{\star}^2 - R_{\rm max}^2 + (R_\text{max}^2 - R_P^2)\langle T \rangle _\lambda }
    {R_{\star}^2 - R_P^2}.
    \label{eq:depth}
\end{align}
Note that Eqs.~\ref{eq:depth} and~\ref{eq:trans} coincide if the atmosphere covers the entire stellar disk. The quantity $(1-T_\lambda)$, measured over the nominal, {planet-frame} absorption line(s) wavelengths, defines the absorption depth of the metastable helium transit. For all practical purposes, this is equivalent to the absorption depth in the 10,830.34$+$10,830.25 \r{A} lines (i.e, the doublet absorption depth).  

Instrumental line broadening and/or broadening due to the planet intrinsic rotation can also be accounted for by the TPM. The former is implemented via a 
simple convolution with a Gaussian of full width half maximum $\FWHM = \bar{\lambda}/R$, where $\bar{\lambda} = 10,830$~\r{A} and $R$ is the instrument resolution. For the latter, the TPM assumes that the planet is tidally locked, with orbital frequency $\omega_p$. The convolution is performed with a Gaussian of $\FWHM = R_\text{eff} \omega_p \bar{\lambda}/c$, where the planet effective radius $R_\text{eff}=R_P\,\sqrt{h/\delta +1}$ is set by a combination of the line absorption depth, $h$, (possibly, but not necessarily, after the LSF convolution), and the optical transit depth, $\delta$. 

Note that in all simulations presented in the next sections we do not include any instrumental or rotational line broadening. 

\begin{figure}
    \center
    \includegraphics[width=1.0\columnwidth]{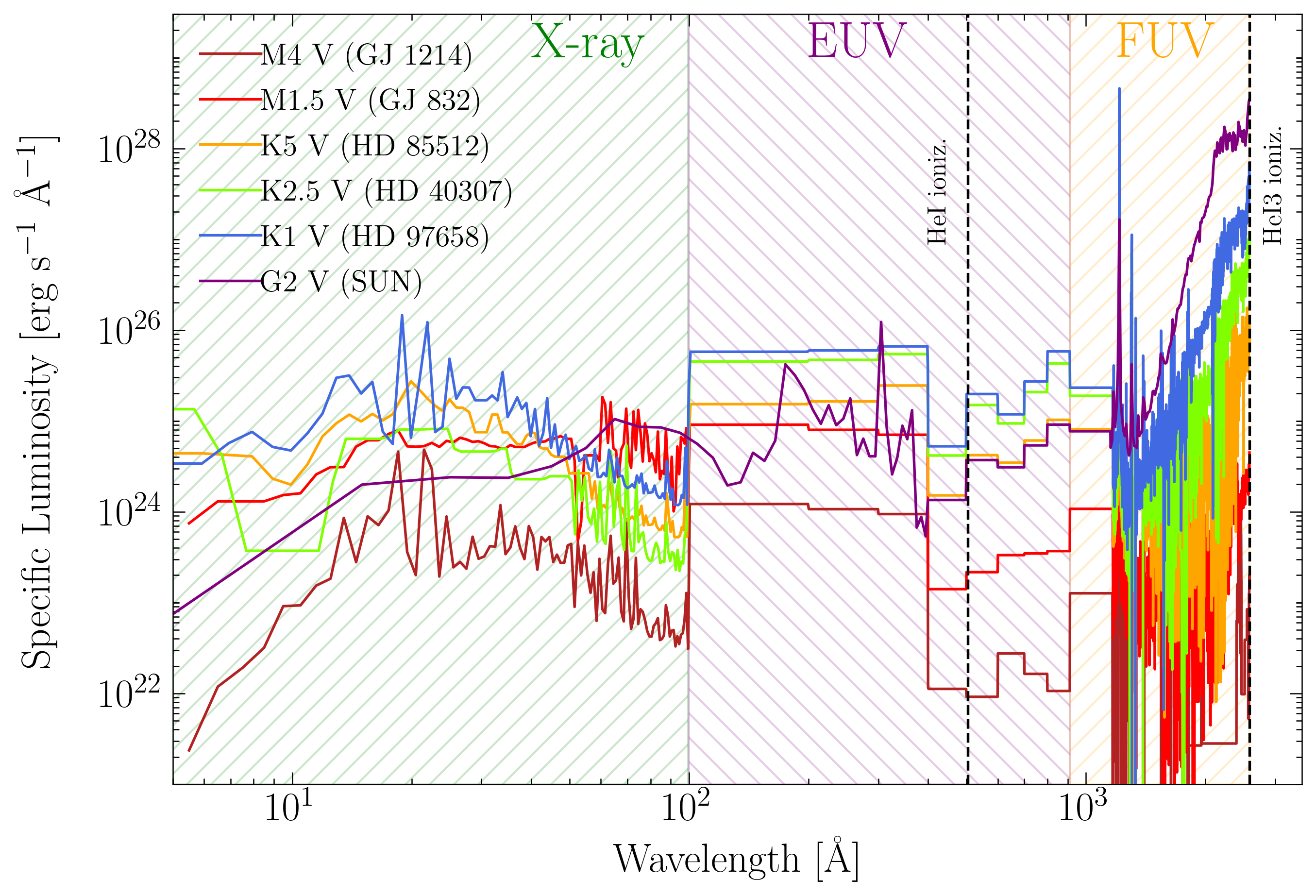}
     \caption{Broadband spectra of six prototypical planet-hosting stars considered in this work. }
     \label{fig:spectra}
\end{figure}

\begin{table*}
\centering
\caption{Stellar parameters for the spectra shown in Figure \ref{fig:spectra}.}  \label{tab:stars}
\renewcommand{\arraystretch}{1.3}
	\begin{tabular}{lllllll}
    	\hline
    	\hline
        Star                                & %
    	Type                                & %
    	$\log L_{\rm XUV}^{(1)}$            & %
    	$\log L_{\rm FUV}^{(2)}$            & %
    	HR$^{(3)}$  	                    & %
        M$_{*}^{(4)}$                       & %
        R$_{*}^{(5)}$                       \\       
    	\cline{1-7}
        GJ 1214       & M4 V   & 26.56 & 26.70 & 0.72                 & 0.18$^{(a)}$  & 0.21$^{(a)}$ \\
        GJ 832        & M1.5 V & 27.49 & 27.60 & 0.76                 & 0.44$^{(c)}$  & 0.44$^{(c)}$ \\
        HD 85512      & K5 V   & 27.94 & 28.60 & 0.22                 &  0.69$^{(c)}$ & 0.69$^{(c)}$ \\
        HD 40307	  & K2.5 V & 28.39 & 29.25 & 0.14                 & 0.79$^{(c)}$  & 0.72$^{(c)}$ \\
        HD 97658	  & K1 V   & 28.51 & 29.85 & 0.046                & 0.85$^{(b)}$  & 0.73$^{(b)}$ \\
        Sun	          & G2 V   & 27.80 & 31.34 & 2.89$\times 10^{-4}$ & 1 & 1 \\
    	\hline
    	\hline
	\end{tabular}
	\tablefoot{(1): logarithm of ionizing ($h\nu\geq 13.6$ eV) luminosity in erg/s. (2):  logarithm of \HeITR~ionizing ($4.8\leq h\nu <13.6$ eV) luminosity in erg/s. (3): hardness ratio $L_{\rm XUV}/L_{\rm FUV}$. (4) Stellar mass in M$_{\odot}$. (5) Stellar radius in R$_{\odot}$. (a) \citet{Cloutier_2021}; (b) \citet{Ellis_2021}; (c) \citet{Stassun_2019}.}
\end{table*}

\section{Absorption Depth as a Function of Star-Planet Parameters}
\label{sec:results} 

We make use of the updated code, combined with the TPM, to assess how different stellar, planetary, and atmospheric parameters affect the physics and observability of metastable helium transits. 
We start with exploring different combinations of measurable system parameters, namely, host stellar type, planet orbital distance, and (unlike previous studies) planet gravity.  

The atmospheric outflow properties depend strongly on the planet gravitational potential, (hereafter defined as $\phi_P=GM_P/R_P$, where $M_P$ and $R_P$ are the planet mass and radius). 
As first shown by \cite{Salz2016b}, the evaporation efficiency plummets for planets with $\log(\phi_P)\gtrsim 13$ (in c.g.s. units). \cite{Caldiroli_2022} demonstrated that this threshold is set by an energetic balance: for high-gravity planets, most of the stellar energy absorption occurs within a region where the average kinetic energy acquired by the ions through photo-electron collisions is insufficient for escape.
As a result, only (mildly-irradiated) low-gravity planets, with $\log(\phi_P)\lesssim 13$, can exhibit {``energy-limited"} outflows, whereby all of the absorbed stellar radiation is used up to drive an outflow. This range encompasses super-Earths and mini-Neptunes, all the way to Neptune-sized planets. \\

These considerations lead us to consider/simulate two (known) planets, each representing a prototypical low-gravity and high-gravity case.
As a low-gravity case, we select HAT-P-11b; a Neptune-sized planet with $M_p=0.081\pm0.009 \, M_J$ and $R_P=0.422\pm0.014 \, R_J$, i.e., $\log(\phi_P)=12.54$. HAT-P-11b is in a $4.88$ day orbit (with semi-major axis $a = 0.0530^{+0.0002}_{-0.0008}$ AU) around a K4 V star with mass $M_{\star} = 0.81_{-0.03}^{+0.02} \, M_{\odot}$ and $R_{\star} = 0.75 \pm 0.02 \, R_{\odot}$ \citep{Bakos_2010}.  For the high-gravity case, we choose HD~189733b, a Jupiter-sized planet with $M_p=1.123\pm0.045 \, M_J$, $R_P=1.138\pm0.027 \, R_J$, yielding $\log(\phi_P)=13.25$.
HD~189733b is in a $2.2$ day orbit ($a=0.03100_{-0.00061}^{+0.00059}$ AU; \citealt{Bonomo_2017}) around a K2 V star with $M_{\star}=0.806\pm0.048 \, M_{\odot}$ and $R_{\star}=0.756\pm0.018 \, R_{\odot}$ \citep{Paredes_2021}.\\ 

Input stellar spectra are drawn from the {\fontfamily{pcr}\selectfont MUSCLES Treasury Surveys v22} \citep{France_2016, Youngblood_2016, Loyd_2016} for the following stars/spectral types: GJ 1214 \citep[M4 V spectral type,][]{Cloutier_2021}, GJ 832 \citep[M1.5 V spectral type,][]{Bailey_2009}, HD 85512 \citep[K5 V spectral type,][]{Pepe_2011}, HD 40307 \citep[K2.5 V spectral type,][]{Tuomi_2013}, and HD 97658 \citep[K1 V spectral type,][]{Ellis_2021}. For the purpose of modeling the EUV portion of the stellar spectra, MUSCLES makes use of empirical scaling relation based on Ly$\alpha$  fluxes \citep{Linsky_2014} and/or differential emission measure models \citep{Duvvuri_2021}. The visible-IR spectra are from PHOENIX atmosphere models \citep{Husser_2013,Allard_2016} 
In addition, we consider the Sun (G2 V spectral type), whose spectrum is taken from SORCE-LISIRD\footnote{\url{http://lasp.colorado.edu/lisird/}}; the 400-1150 \r{A} interval is also based on the empirical relation with Ly$\alpha$ \citep{Linsky_2014}. \\

Shown in Figure~\ref{fig:spectra} are the broadband spectra of the six chosen spectral types. Whereas \Hy-ionizing photons are the main driver of the outflow thermodynamics, it is the \HeI~ionizing photons that are mainly responsible for populating the \mhe\ state (via recombinations), while FUV photons deplete it, via photoionization (electron collisions also contribute to the depletion).
As a result, {the density profile of the metastable helium state is driven by a combination of the intensity of the \HI~ionizing flux at the planet's surface, {and} the XUV to FUV hardness ratio}, defined as ${\rm HR}=L_{\rm XUV}$/$L_{\rm FUV}$ (see Table~\ref{tab:stars} for a list of HR values for the chosen spectral types). 

In addition, as we demonstrate below, the transit signal is very sensitive to the atmosphere's physical extent compared to the stellar disk size; this ratio can be close to unity for planets orbiting M-type stars at moderate orbital distances. 
\begin{figure*}
\center
\includegraphics[width=1.0\textwidth]{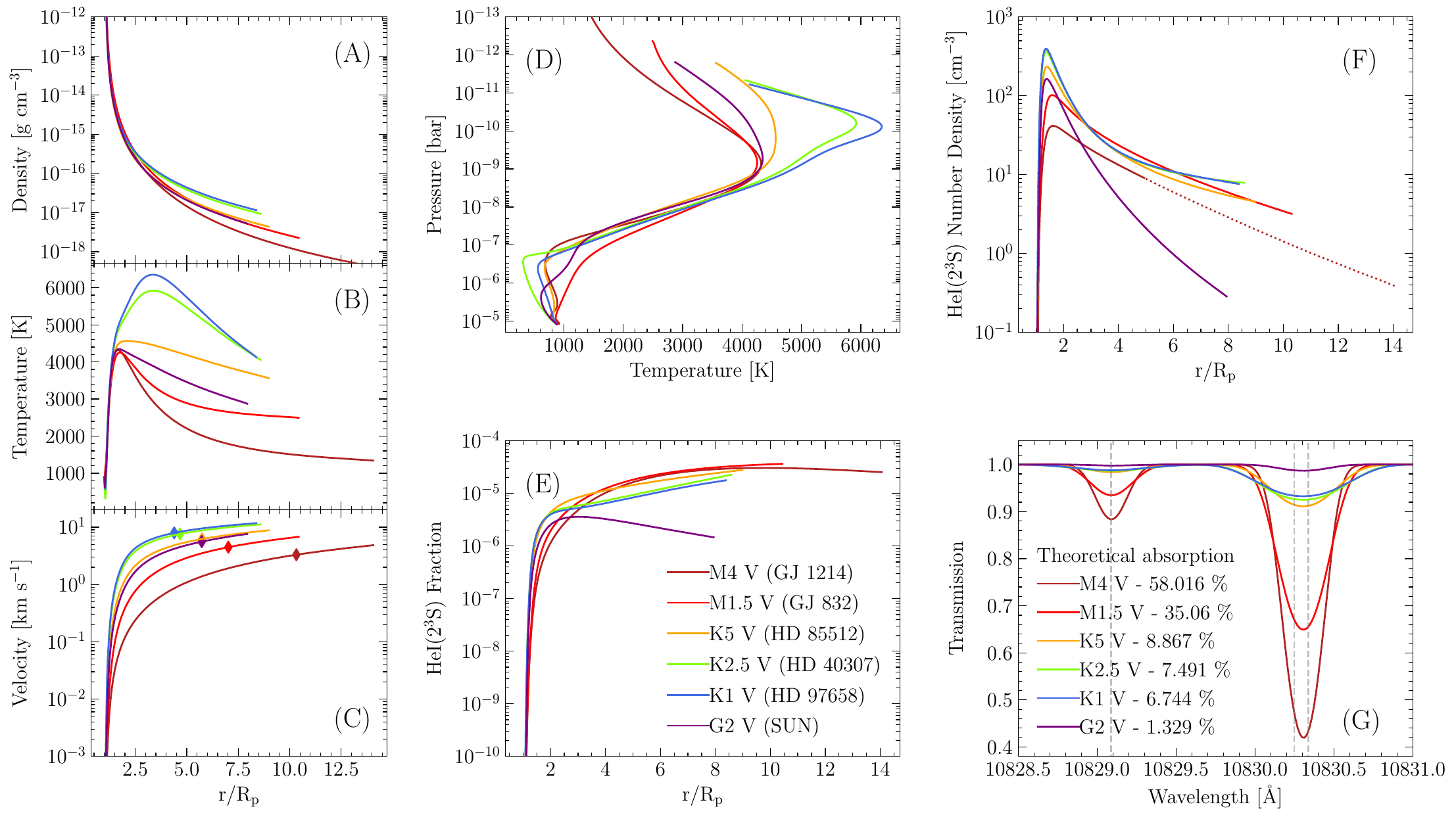}
     \caption{ATES simulations of the Neptune-sized planet HAT-P-11b, subject to varying stellar irradiation fields. The planet orbital distance is fixed, to 0.053 AU. The different colors correspond to the host spectral types shown in Figure \ref{fig:spectra}.  Shown are the radial profiles of the outflow total density (A); temperature (B); velocity (C), where a diamond symbol marks the location where the outflow becomes supersonic; \He~fraction in the triplet state (E); \mhe\ number density (F). All profiles extend from the planet surface to the system's Hill radius. For the M4 V case, the Hill radius exceeds the size of the stellar disk (past which the outflow profile is shown as a dotted line in panel F). Panel D shows the outflow pressure-temperature diagram; the resulting \mhe\ transmission probability is shown in panel G. }
     \label{fig:HATP11_types}
\end{figure*}
\begin{figure*}
\center
\includegraphics[width=1.0\textwidth]{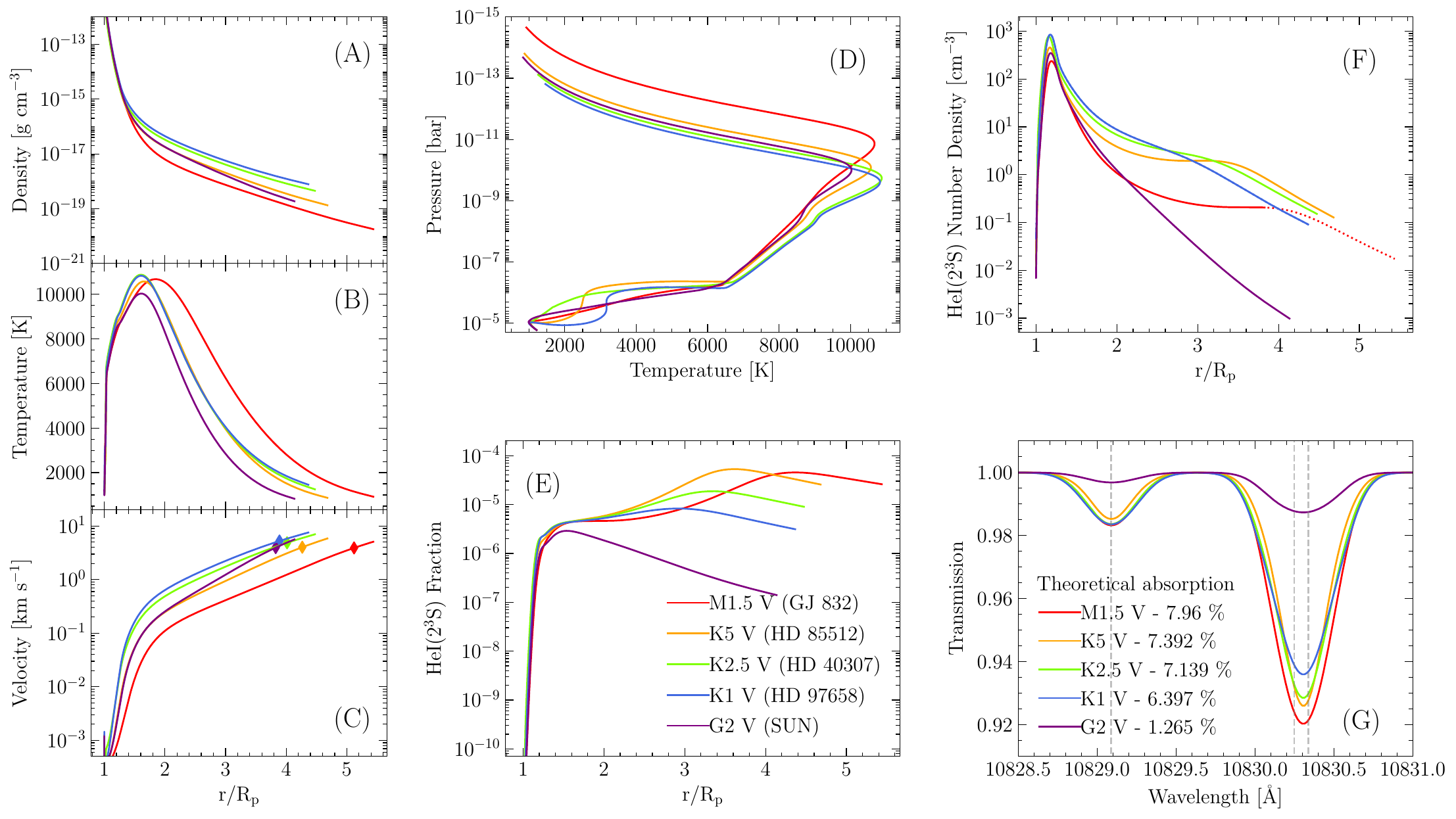}
     \caption{Same as Figure~\ref{fig:HATP11_types}, but for the gas giant HD 189733b, at an orbital distance of 0.031 AU.}
     \label{fig:HD189733_types}
\end{figure*}

\subsection{Varying Host Spectral Type}

Here we present and discuss a suite of ATES simulations where the Neptune and Jupiter-sized planets HAT-P-11b and HD 189733b -- placed at their inferred orbital distances\footnote{As a result, in all cases HD 189733b's irradiance is about 3 times as high as HAT-P-11b's for the same choice of host star.} -- are exposed to the incident stellar spectra shown in Figure \ref{fig:spectra}. Throughout, the He-to-H abundance is set to the cosmological value 1/12. 

\textit{\subsubsection{Neptune-sized planets}}
Figure~\ref{fig:HATP11_types} illustrates the results for the case of the Neptune-sized planet HAT-P-11b. As expected, the host stars with higher XUV fluxes yield higher outflow mass density, temperature, and velocity (panels A, B and C, respectively). The inferred, steady-state mass-loss rates (listed in Table~\ref{tab:trans}) range between $\simeq 10^{9}$ g s$^{-1}$ (for the M4 host) and $\simeq 5\times 10^{10}$ g s$^{-1}$ (K1 host). 
In all cases, the outflow becomes supersonic within the simulation domain (i.e., within the Hill radius).

A key result (panel B) is that the outflow is not well approximated by an isothermal Parker wind. An isothermal profile is only acceptable for M-type spectra at very large distances from the planet surface ($r\gtrsim 5 R_P$). The temperature profiles exhibit sharp inversions, as also shown by the pressure-temperature diagrams (panel D). 
The temperature profiles begin to rise very sharply starting from the height where the bulk of the stellar radiation is absorbed, at pressures in the range $10^{-7}-10^{-5}$ bar. 
This occurs at about 1.1 $R_P$ in the case of HAT-P-11b, which has a relatively high atmospheric density. \textit{The second temperature inversion occurs further out, at pressures lower than $10^{-9}$ bar (and/or heights above 1.5 $R_P$); the location where the total cooling rate starts to decrease corresponds to the beginning of the temperature decline.}
\begin{table}
\centering
\caption{Simulated mass loss rate and \mhe\ transits.}    \label{tab:trans}
    \renewcommand{\arraystretch}{1.2}
	\begin{tabular}{lcccc}
    	\hline
    	\hline
         ~                         & %
     \multicolumn{2}{c}{HAT-P-11b} & %
     \multicolumn{2}{c}{HD189733b}\\        
         Type& 
         $\log \dot M^{(1)}$& 
         Abs\%$^{(2)}$& 
         $\log \dot M^{(1)}$& 
         Abs\%$^{(2)}$ \\  
    	\cline{1-5}
        M4 V      & 9.03 & 58.0 & -- & --  \\
        M1.5 V       & 9.96 & 35.1 & 8.04 & 7.9 \\
        K5 V     & 10.23 & 8.9 & 8.84 & 7.4 \\
        K2.5 V	 & 10.62 & 7.5 & 9.40 & 7.1 \\
        K1 V	 & 10.72 & 6.7 & 9.65 & 6.4\\
        G2 V         & 10.14 & 1.3 & 8.86 & 1.3\\
    	\hline
    	\hline
	\end{tabular}
	\tablefoot{(1): Logarithm of the mass loss rate, in g s$^{-1}$. (2): Absorption depth in the doublet core.}
\end{table}

Panels F and E show the radial profiles of the \HeITR~number density and triplet fraction; the ensuing \mhe\ transmission signals are shown in panel F. The main takeaway is that these do not depend on the host spectral type in a straightforward way. 
\HeITR~is mainly populated by \HeII~recombinations, and depleted by collisions with electrons and/or FUV photons. The lowest \mhe\ density profile is yielded by the M4-type and G2-type spectra, albeit for different reasons. 
In the former case, this is caused by the low \HeI~ionization rate, which in turn implies a low recombination rate. In the latter case, the low density arises from intense FUV .

It is interesting to note that in all cases but the Sun (G2 V), the \HeITR~density profile exhibits a shallower decline compared to the total density profile, implying that the outermost regions of the atmosphere are responsible for a relatively large fraction of the absorption signal. 
This behavior is driven by the main mechanisms that populate/deplete the metastable state, and can be understood by inspecting the radial profiles of the reaction rates, shown in Figures \ref{fig:a3} in the Appendix. At large distances from the planet surface, the temperature decrease makes the balance between recombinations and electron collisions shift toward a (relative) overabundance of \HeI~atoms in the metastable state. 
We also note that the \HeITR~fraction values in the outermost part of the flow are in good agreement with the analytical estimate in Sect. \ref{sec:ATES_updates} (i.e., \HeITR/\He$\sim 10^{-5}$).  
The sharp decline in the \HeITR~density profile for the case of the Sun occurs because of its intense \HeITR-ionizing FUV flux which efficiently depletes the metastable state via photoionization.\\

The resulting, wavelength-dependent transmissions $T_\lambda$ are shown in panel G. The transit absorption depths listed in Table~\ref{tab:trans} correspond to the value $(1-T_\lambda)$ at 10,830 \r{A}. 
The most notable result is that M-type hosts yield the highest absorption depths, as high as $\sim 35[/60]$\% for the case of a M1.5[/4] V host. This is in contrast with the results by \cite{Oklopcic_2019}, which concluded that K-type hosts can be expected the strongest absorption signals. 
This discrepancy, we argue, is largely driven by the low outflow temperatures (2,000-3,000 K), which in turn imply much higher $\alpha_3$ rates (Figure~\ref{fig:rec_ratio}) than those given by the 7,500 K chosen by \cite{Oklopcic_2019} for the M-type case (see their figure 7). 

Additionally, we emphasize that higher metastable state density profiles do not necessary imply higher absorption depths. This is partly because the absorption depth depends on the ratio $(R_{\rm max}/R_\star)^2$, i.e., the fraction of the stellar disk that is actually eclipsed by the planet's outflowing atmosphere. The small radii of M-type stars, combined with the large system Hill radii, tend to yield very deep absorption signals, up to $\simeq 60\%$ for the case of an M-type host (for this specific planet gravity and orbital distance).
We caution, however, that the expected absorption depth depends somewhat counter-intuitively on orbital distance; this is further discussed in Sect. \ref{subsec:distance}. 

Because of the larger stellar radii, K-type stars produce much shallower absorption depths, of order 7\%-10\%. Interestingly, the lowest absorption depth (amongst K-types) is expected for the hottest type (K1 V). This is because thermal and bulk motions-induced Doppler broadening cause the line to be broader, and hence shallower in the core. 
Finally, due to the large stellar radius and low \HeITR~density profiles, a Sun-like host is expected to yield a very modest absorption features, $\lesssim  1.5\%$.

\begin{figure*}
\center
\includegraphics[width=1\textwidth]{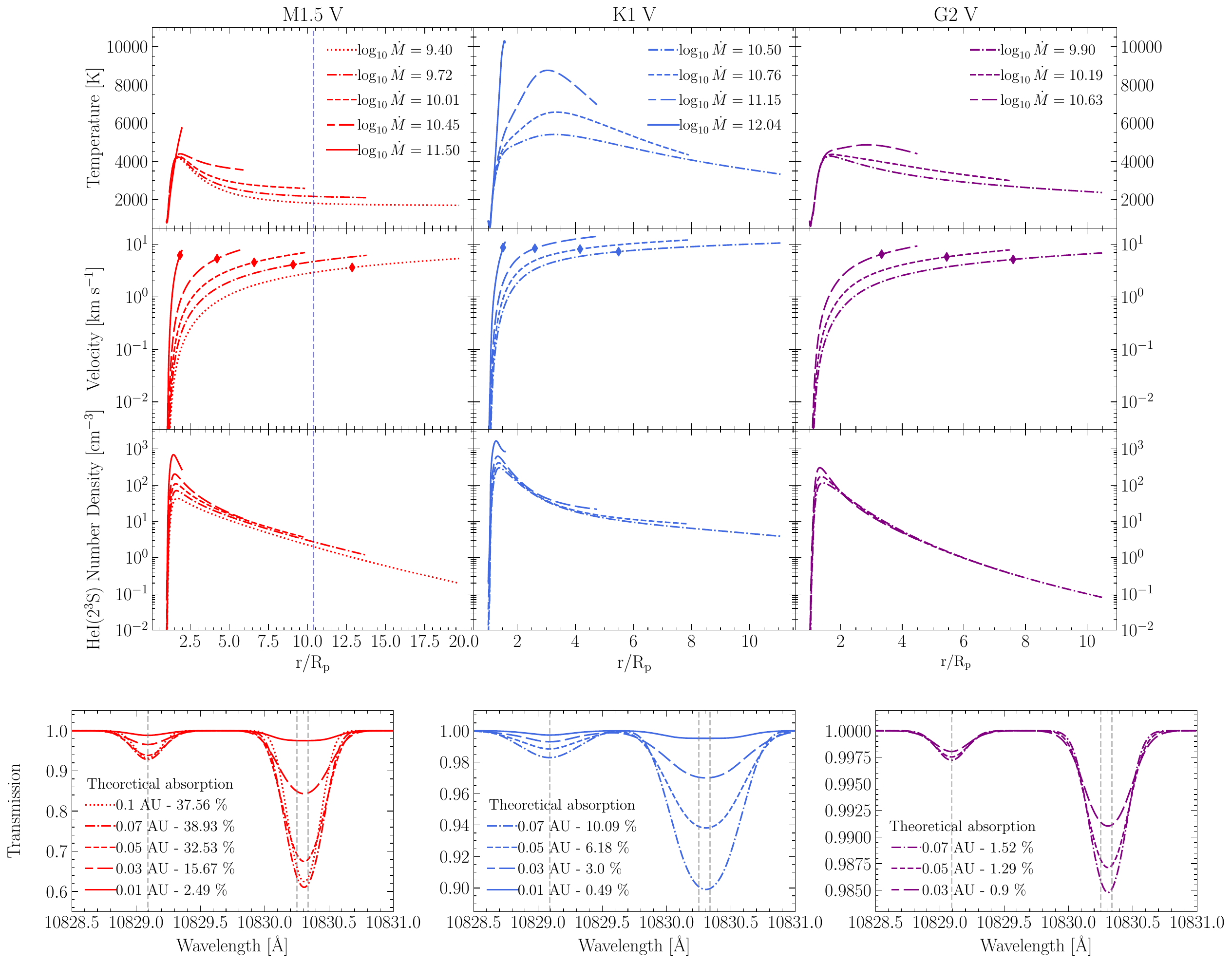}
\caption{Simulated outflows from the Neptune-size planet HAT-P-11b, orbiting a M1.5 V (left column), K1 V (middle column), and G2 V (right column) type star at orbital distances in the range 0.01-0.1 AU. The top, second and third row show the outflow temperature, velocity and \HeITR~density profiles. The dashed vertical line shown for the M1.5 V case indicates the extent of the stellar radius (only for this spectral type can the planet atmosphere be more extended than the actual host star). The diamond symbols in the velocity profiles mark the location where the outflow becomes supersonic. The bottom row shows the resulting \mhe\ transmission probability.}
     \label{fig:HATdistance}
\end{figure*}
\begin{figure*}
\center
\includegraphics[width=1\textwidth]{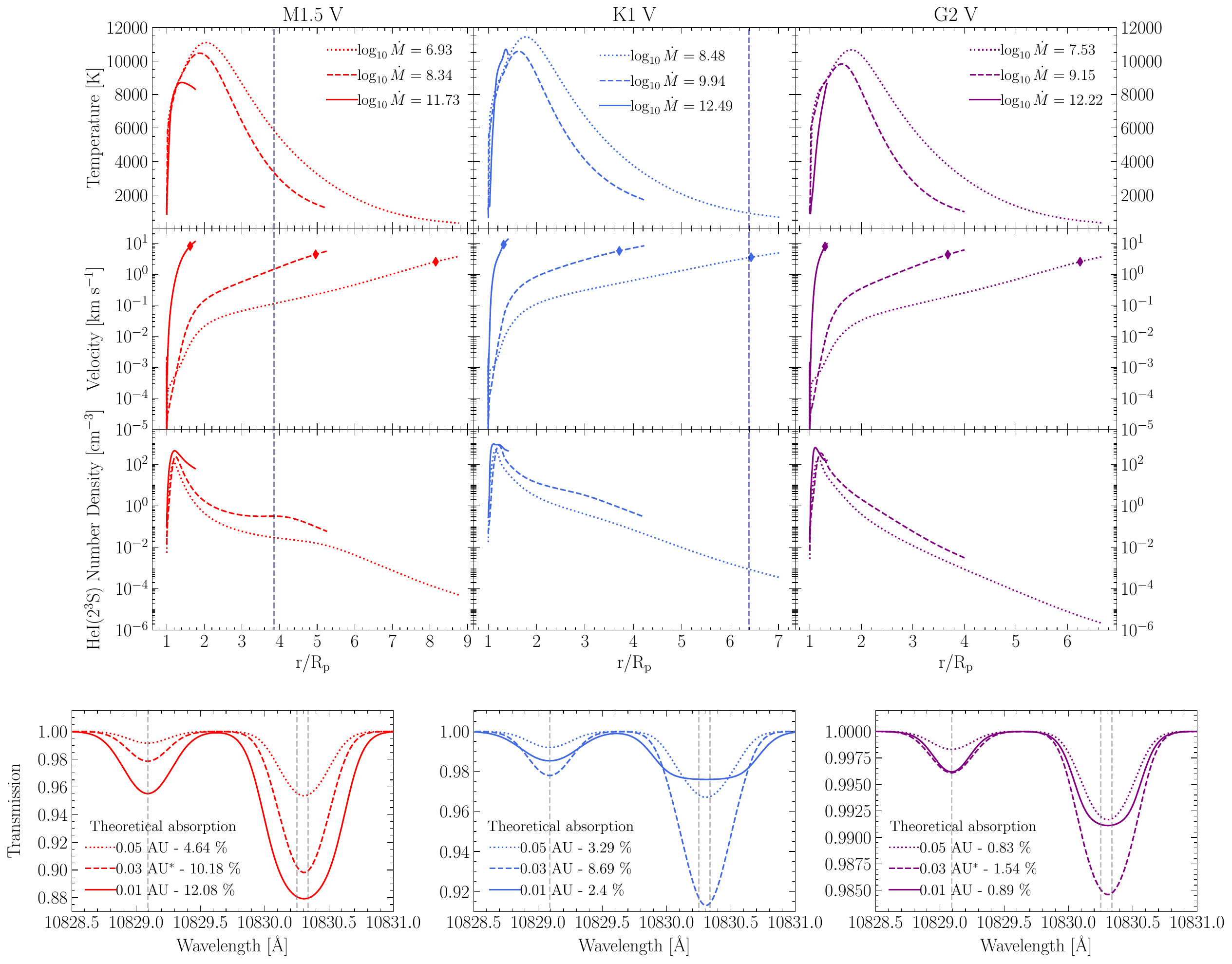}
     \caption{Same as Figure~\ref{fig:HATdistance} but for the gas giant HD 189733 b. Owing to the planet's strong gravity, the outflow fails to converge for an orbital separation larger than 0.05 AU, so only 3 separations are considered in this case.}
     \label{fig:HDdistance}
\end{figure*}

\textit{\subsubsection{Jupiter-sized planets}}

Figure~\ref{fig:HD189733_types} shows the results of varying the host spectral type for the case of the Jupiter-sized planet HD 189733b (at 0.031 AU). As expected, in spite of the smaller orbital distance, this gas giant yields much lower mass outflow rates -- between $\simeq 10^{8}$ g s$^{-1}$ (M1.5 V-type) and $\simeq 5\times 10^9$ g s$^{-1}$ (K1 V; see Table~\ref{tab:trans}). In fact, the code fails to converge for the case of a M4 V spectral type, where the outflow likely approaches Jeans escape. 

Lower $\dot M$ values are driven by the lower density and velocity profiles (panel A and C) compared to the case of HAT-P-11b. Also, the temperature profiles are fairly insensitive to the stellar spectral type (panel B), and never resemble an isothermal wind. All temperature profiles start to increase right at the planet radius, reaching a peak value of $T\simeq$ 10,000\,K close to $2 R_P$, and then roll off quite steeply toward 
$T \simeq$1,000-2,000\,K, at the Hill radius (please note that, in this cases, temperature values very close to the planet surface may be impacted by numerical accuracy; as a result, the $P$-$T$ curves around $\simeq 10^{-5}$\,bar may be non-physical).  

The \HeITR~fraction and number density profiles (panels E and F) are qualitatively similar to those of HAT-P-11b, although HD 189733b's \HeITR~fraction profiles exhibit a ``saddle" in the outer regions (except for the G2 V spectrum). As shown in Figure \ref{fig:a4}, this trend is caused by a sharp decline in the $q_{31}$ transition rates, accompanied by the concurrent rise in the photoionization rate (owing to the steady decline in the photoionization rate, this does not happen in the case of a G2 V spectrum).
Since the outflow velocities are always modest (below $10$\,km/s), Doppler broadening is dominated by random thermal motion. Combined with the fact that the outflow temperatures (both peak and profile) are nearly independent of spectral type, this yields comparable line broadening for different spectral types. For K and G-types, the absorption depth is driven by the \HeITR\ density, with M-types standing out again thanks to the relatively small stellar disk size compared to the atmosphere's (see Table~\ref{tab:trans}). \\

\begin{figure*}
    \centering
    \includegraphics[width=12cm]{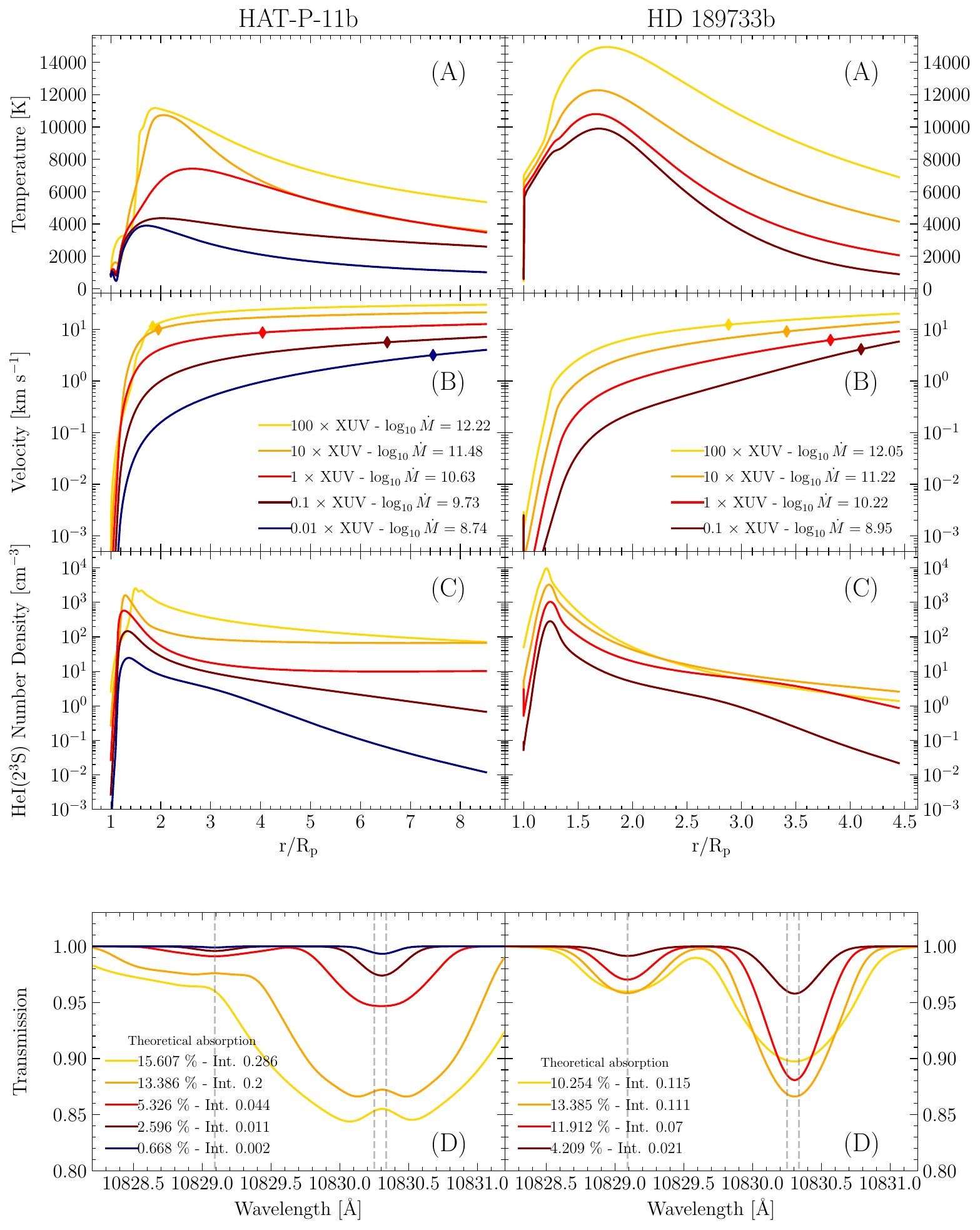}
    \caption{Effects of different stellar XUV on HAT-P-11b and HD 189733b. Left column: HAT-P-11b, irradiated by the K4 V star HAT-P-11, with varying XUV intensity. Right column: HD 189733b, irradiated by the K2 V star $\varepsilon$~Eri, with varying XUV intensity. Temperature, velocity and \HeITR~number density profiles are shown in panel A, B, and C, respectively. Panel D shows the resulting transmission.}
    \label{fig:vary_XUV}
\end{figure*}

\subsection{Varying Orbital Distance}
\label{subsec:distance}
  
Next, we vary the orbital distances of the two planets under scrutiny, using three host spectral types (M1.5 V, K1 V, and G2 V). The minimum distance is set to 0.01 for all cases; the maximum distance d is set by the code convergence requirements.
Figures~\ref{fig:HATdistance} and \ref{fig:HDdistance} show the results of this investigation for the Neptune and Jupiter-sized planet, respectively\footnote{Please note that the results obtained for HAT-P-11b[HD 189733b] at an orbital distance of 0.05[0.03] AU are essentially the same as discussed in the previous Section, since the planet orbital distance is 0.053[0.031] AU}.

For a given host spectral type, orbital distance affects the metastable helium absorption signal in two main ways: first, a more intense flux triggers a denser outflow (and higher mass outflow rate). Second, the location of the Hill radius (and thus the nominal extent of the atmosphere compared to the stellar disk's) is affected by the orbital distance. The two effects act in opposite directions: which dominates, depends on the specific case under consideration.

Starting with the Neptune-sized HAT-P-11b, irradiated by an M1.5 V star (left column in Figure~\ref{fig:HATdistance}), the first thing to note is that whereas $\dot M$ increases with decreasing orbital distance, the corresponding \mhe\ absorption depth falls from $\simeq 38\%$ to $\simeq 3\%$ for closer orbits. This somewhat counter-intuitive effect is caused by the much reduced extension of the planet's outflow compared to the stellar disk. It is worth noting that, at a distance of 0.01 AU, the expected absorption depth for this choice of parameters approaches its geometric maximum, i.e. $(R_H^2-R_P^2)/(R_\star^2-R_P^2)$. This, together with the apparent saturation in the line Doppler wings, indicates that the whole outflow is optically thick to \HeITR, yielding a shallow, optically thick absorption feature. The trends are qualitatively similar for the K1 V and G2 V spectral types; shallower, optically thick absorption features can be expected at closer distances. Nevertheless, the larger radii of K-types result in shallower absorption depths compared to G-types. \\

The results are qualitatively different for the Jupiter-sized planet HD 189733b (see Figure~\ref{fig:HDdistance}), mainly because of the much lower mass-loss rates.  As for HAT-P-11b, irradiation by a M1.5 V host at 0.01 AU produces an absorption feature that is close to its geometric maximum (though the profile is clearly still not saturated). However, larger orbital distances yield such low mass-loss rates ($\lesssim 2\times 10^8$ g/s) that the resulting absorption depths are shallower in spite of the larger atmospheres.
For K1 V and G2 V irradiation (middle and right columns in Figure~\ref{fig:HDdistance}) at 0.01 AU and 0.03 AU we observe similar trends as for HAT-P-11b, i.e., closer orbits lead to shallower absorption depth. At 0.05 AU, the trend is reversed owing to insufficient EUV irradiance.

To summarize, the extent of the atmosphere (that is, the size of the Hill radius) is the dominating parameter in setting the expected \mhe\ absorption depth for a low-gravity, Neptune-sized planet such as HAT-P-11b. The feature becomes deeper for larger orbital distances (at least up to 0.05 AU). We caution that this result stems directly from the choice of the Hill radius as the integration limit; we come back to this point in Sect.  \ref{sec:conclusions}.

For gas giants, such as the Jupiter-sized HD 189733b, the same trend is observed only for distances closer than 0.03 AU off a K1 V and/or G2 V type host. Further out, the absorption depth decreases with distances. For M1.5 V hosts, the planet gravity wins, in the sense that the absorption depth decreases with increasing orbital distance in spite of any geometrical effects. 

Once more, we stress how the expected \mhe\ densities (and corresponding transits) are highly sensitive to the outflow temperature profiles, and that an isothermal approximation is likely to fail by more than 1 order of magnitude (see \citealt{vissa22} for a discussion on the applicability of Parker wind solutions, and \citealt{linssen22}, on how the Cloudy photoionization code can be used to resolve the temperature-mass loss rate degeneracy that affects Parker wind models).

\subsection{Varying XUV and He abundance}
\label{subsec:He}

We investigate how the well known uncertainties in the (shape and intensity of the) stellar XUV spectrum affect the physics and observability of the metastable helium feature. At variance with \cite{Oklopcic_2019}, here we only vary the XUV portion of the spectrum, since (unlike the EUV) the stellar FUV flux can be directly measured (we verified that, as expected, progressively increasing the FUV flux decreases the metastable helium number density, and thus the expected absorption depth; the effects of varying the FUV portion of the spectrum on the mass loss rates are negligible).

In an effort to minimize the number of variables, we simulate the case of HAT-P-11b irradiated by its actual host star, i.e., a K4 V type; we make use of the spectrum compiled by \cite{Ben-Jaffel_2022}, which uses HST and XMM-Newton data for the FUV and X-ray portion, respectively, and reconstruct the EUV spectrum with a coronal and transition region model. 
We then modify HAT-P-11's spectrum by rescaling the XUV spectrum by a factor $\times$$100, 10, 0.1, 0.01$.

As shown in the left column of Figure~\ref{fig:vary_XUV}, 
varying the XUV portion of the spectrum yields drastic changes in the metastable helium transits, as also found by \cite{Oklopcic_2019}. Higher XUV fluxes naturally drive higher density outflows, as well as higher \HeITR~density profiles (panel C). The $\times$10 and $\times 100$ XUV cases correspond to very high temperatures and velocities across the whole outflow (panel A and B); these give rise to very broad absorption features, characterised by deep absorption wings. \\

For the gas giant HD 189733b we adopt the spectrum of the K2 V star $\varepsilon$~Eri (from MUSCLES), which is routinely used in the literature as a proxy for HD 189733's spectrum, and re-scale the XUV portion by a factor $\times$$100, 10, 0.1$ (the outflow fails to converge for a factor 100 times lower XUV flux, owing to the high planet gravity).  As shown in the right column of Figure~\ref{fig:vary_XUV}, the results are qualitatively different from those obtained for HAT-P-11b. In this case the temperature profiles are generally high enough to make the \HeITR~abundance fairly stable against temperature differences of a factor $\lesssim 2$ (see Figure~\ref{fig:rec_ratio}). Combined with the relatively low outflow velocities ($\lesssim 20$ km/s) this yields a modest dependence of the \HeITR~density profiles on the intensity of the ionizing radiation, particularly in the outer outflow regions, where He is single or even doubly ionized. As a result, the absorption depths are fairly independent on the XUV intensity; only in the $100\times$XUV case yields somewhat lower absorption, mainly because the feature becomes broader (and thus shallower) owing to the larger flow bulk velocity.

Finally, we examine the (critical) role of the chosen helium-to-hydrogen abundance. We repeat the same simulations as above, and vary the He/H number density ratio in the range $10^{-2}$ to 1. This is supposed to bracket the broad range of values that have been inferred in the literature to reproduce the observational data, particularly as it pertains to non detections. 

\begin{figure}
    \centering
    \includegraphics[width = .8\columnwidth]{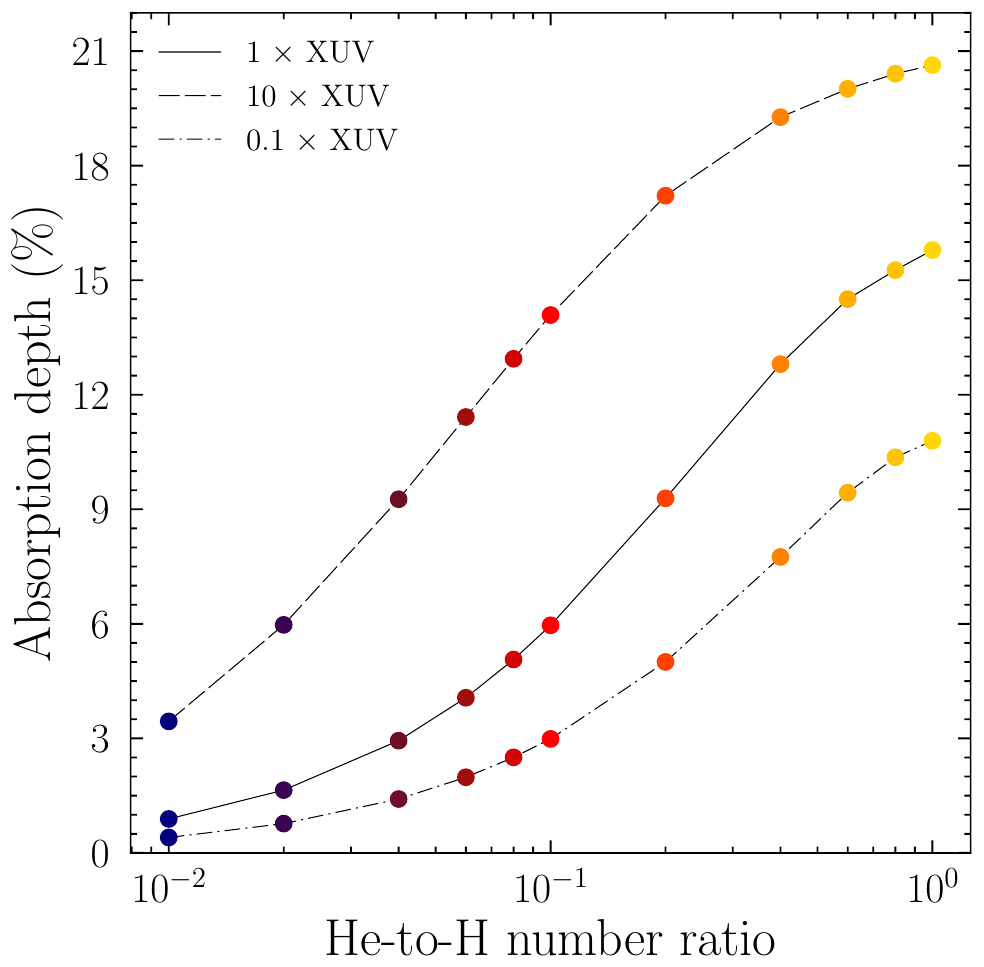}
    \caption{Absorption depth as a function of the chosen He/H number density ratio for the case of HAT-P-11b, irradiated by HAT-P-11 ($1\times$XUV), as well as an HAT-P-11-like spectrum with 10 times lower and higher XUV irradiance. 
    }
    \label{fig:varyHeH}
\end{figure}

The resulting \mhe\ absorption depths are shown in Figure~\ref{fig:varyHeH}. 
The expected transit depth varies between $\simeq 1\%$ for He/H=0.01, up to $\simeq 16\%$ for He/H=1. However, this quantity is highly degenerate with the chosen stellar XUV intensity, such that the same \mhe\ absorption depth -- e.g. $\sim$10\% -- could be attained with $0.1\times$XUV and He/H=0.5, baseline XUV and He/H=0.2, or $10\times$XUV and He/H=0.04. 

Observations of HAT-P-11b, with CARMENES \citep{Allart_2018}, measured an absorption depth at 10,830 \r{A} of $(1.08 \pm 0.05) \%$. Based on our simulations, and also accounting for instrumental resolution and planet rotation, this level can be achieved with a baseline (i.e., $1\times$) XUV spectrum and He abundance of $\sim$0.01, or with a $0.1\times$ XUV spectrum and He abundance of $\sim$0.03. Similar conclusions are reached by \citet{Allart_2018} using a 3D model and a somewhat different input spectrum from \citet{Ben-Jaffel_2022}. 

For the case of HD 189733b, different \HeITR~absorption depths have been reported in the literature: $(0.88\pm 0.08)\%$, with CARMENES \citep{Salz_2018}, $(0.75\pm 0.03)\%$, with GIANO-B \citep{Guilluy_2020}, and $(0.420\pm 0.013)\%$ with Keck/NIRSPEC \citep{Zhang2022},  possibly indicative of variable absorption \citep{Guilluy_2020,Zhang2022,Pillitteri_2022}. Again, in order to match such low values our simulations require a low XUV flux (10\% of the nominal $\varepsilon$~Eri irradiation), combined with an extremely low He/H number ratio, $\lesssim 8\times 10^{-3}$. Qualitatively similar conclusions were drawn by \citet{Zhang2022}, using TPCI \citep{Salz2015}.

\section{Conclusions}
\label{sec:conclusions}

We make use of the public 1D photoionization hydrodynamic code ATES \citep{Caldiroli_2021} to model atmospheric absorption by metastable helium during planetary transits. Since it does not suffer from massive interstellar scattering, \mhe\ should provide a more direct probe of atmospheric escape compared to Ly$\alpha$; unlike Ly$\alpha$, it is also observable from the ground.

In the first part of the Paper, we present new modifications to the main code, aimed at solving the continuity equations separately for the neutral helium triplet and singlet; we then develop a new transmission module which calculates the expected line absorption depth during transits. 
We use the new version of the code and the TPM, which are both publicly available\footnote{\url{https://github.com/AndreaCaldiroli/ATES-Code}}, to study how known planetary parameters affect the expected depth of the metastable helium absorption feature. Our investigation features two main improvements with respect to the foundational work by \cite{Oklopcic_2019}: (i) for a given system, the code calculates the outflow properties (namely: temperature, pressure, ionization fraction, density and velocity profiles, along with the steady-state mass loss rate) concurrently with the \mhe\ absorption depth, as opposed to fixing/assuming suitable values for the temperature and mass outflow rate; (ii) in addition to the role of stellar irradiation (shape and intensity), the dependence of the \mhe\ absorption signal on planetary gravity is also explored (as shown by \citealt{Caldiroli_2022}, the outflow efficiency depends strongly on this parameter, and so does the mass outflow rate). 

Point (i) is critical as it enables to model atmospheric escape self-consistently, thereby bypassing the well known degeneracy between mass outflow rate and temperature (see, e.g., \citealt{linssen22}). Importantly, we show that the outflow is seldom isothermal; since the \mhe\ density profile is extremely sensitive to temperature (Figure~\ref{fig:rec_ratio}), foregoing the isothermal assumption yields non-negligible (qualitative and quantitative) changes in the expected transit depths. \\

At variance with \cite{Oklopcic_2019}, who identified K type stars as favorable hosts, we argue that nearby M dwarfs may be prime targets for metastable helium transmission spectroscopy. More specifically, the case of a Neptune-sized planet orbiting a late type M dwarf between $\sim$0.05-0.1 AU could yield intrinsic (i.e., before instrumental or other broadening) \mhe\ transit depths in excess of 30\%. Such high absorption depths are due to the combination of low outflow temperatures and small geometrical ratios between the planet atmosphere (which is assumed to extend up to the system's Hill radius) and the stellar disk. 
We caution, however, that no such system is currently known. Moreover, such high absorption depths are only expected for the case of a primordial envelope with cosmological helium to hydrogen number density ratio. A further complication arises from strong stellar winds (which we entirely neglect): as shown by, e.g., \cite{Vidotto_2020}, \cite{carolan21}, and \cite{macleod}, the location where the stellar wind's and atmospheric ram pressures reach equilibrium (relative to the sonic point and the Hill radius) likely determines the extent and geometry of the atmospheric outflow. In the most extreme case, where pressure balance is reached at an height (above $R_P$) where the outflow is still sub-sonic, the outflow may be suppressed entirely.

Our analysis also demonstrates that the physics driving the strength of the atmospheric absorption signal during transits is far from straightforward. As an example, the \mhe\ transit depth is not always a decreasing function of orbital distance. This counter-intuitive trend is caused by the reduced (assumed) extension of the atmosphere compared to the stellar disk as the planet moves closer to the host star. 

It is important to note that, while isothermal models can integrate the outflow out to arbitrary distances from the inner boundary, our model is inherently limited to the Hill radius, outside of which the outflow velocity is effectively unknown. However, any unbound material outside of the Hill radius is likely to form a permanent layer of absorbing gas that adds to the transiting exoplanet atmosphere. This extra absorption would be very difficult to isolate with standard setups for spectroscopic transit observations, which typically cover the transit itself plus a relatively short ingress and egress phase\footnote{Whether this material settles along the orbital equator in a torus-like structure, or takes on a cometary shape depends on the intensity of the stellar wind (see, e.g. figure 1 in \citealt{macleod}).}. With this setup, any extra absorption due to a permanent screen of material is likely to be attributed to the star itself. Only in the case of a very long observation that spans the full planetary orbit would this extra absorption become apparent, as shown by the recent (remarkable) result by \cite{Zhang_2023} on HAT-P-32 b, showing evidence for extra absorbing material extending out to some 50 times the planet radius.\\

Whereas metastable helium absorption is, in principle, a more viable and straightforward diagnostic of exoplanet atmospheric escape compared to Ly$\alpha$, the growing body of observational evidence has proven hard to parse. We confirm the strong degeneracy between He abundance and intensity of the XUV radiation field. Generally speaking, low excess absorption (at the few \% level) can be easily reproduced with extremely low He abundances, or, a combination of low EUV irradiation and low He abundances. Similar conclusions have been drawn in the literature for several targets, and using different types of models, from Parker winds to sophisticated 3D models. 

To date, observations of some 40 systems have yielded an even greater number of strict upper limits, below 3-4\%\ (see e.g. \citealt{Bennett_2023} and \citealt{Guilluy_2023}). When compared against the spectroscopy data, our modelling --consistent with literature results-- suggests that either a large fraction of the target systems may have helium depleted envelopes, and/or, that their EUV spectra are systematically overestimated.

In principle, a concurrent fit to the H$\alpha$ and/or Ly$\alpha$ lines, along with the HeI(2$^3$S) triplet, could solve the degeneracy between He/H abundance and EUV flux, by constraining the former. An example of this kind analysis is given by \cite{Yan_2022} for WASP-52 b, where H$\alpha$ and HeI(2$^3$S) absorption have been detected during transits. We note, however, that the steady state number density of neutral hydrogen in the $n=2$ state is set mainly by Ly$\alpha$ excitation from the ground state. In turn, this depends on the intrinsic Ly$\alpha$ profile emitted by the star, and on the (notoriously complex) details of the Ly$\alpha$ radiative transfer, thus making this kind approach highly impractical.

As also suggested by \citet{fossati22}, a viable mechanism for suppressing the helium to hydrogen ratio in the upper atmosphere is phase separation, a process whereby neutral helium separates from liquid metallic hydrogen and ``rains" onto the planet interior  (\citealt{Fortney2004}, and references therein; it is worth noting that the timescale over which this process begins to operate is a strong function of planet mass, with smaller planets, of the order of 0.15 $M_J$, reaching He-to-H mass fractions of $\sim$0.05, or number ratios of $\sim$0.015). Since phase separation is only relevant for (sub-Jovian) planets beyond a few AU, it follows that, if confirmed, very low He abundances in the atmospheres of close-in gaseous planets would argue for the need of efficient migration. 
Observations of exoplanet host stars in the EUV band, e.g. with ESCAPE \citep{France_2022}, will be critical to solving the ongoing debate around helium absorption measurements in exoplanet transits, and their implications for evolutionary models. 

\begin{acknowledgements}
We are grateful to the anonymous reviewer for the insightful comments and suggestions.
\end{acknowledgements}


\bibliographystyle{aa} 
\bibliography{bibliography.bib} 

\begin{thebibliography}{83}
\expandafter\ifx\csname natexlab\endcsname\relax\def\natexlab#1{#1}\fi

\bibitem[{{Allard}(2016)}]{Allard_2016}
{Allard}, F. 2016, in SF2A-2016: Proceedings of the Annual meeting of the
  French Society of Astronomy and Astrophysics, ed. C.~{Reyl{\'e}},
  J.~{Richard}, L.~{Cambr{\'e}sy}, M.~{Deleuil}, E.~{P{\'e}contal},
  L.~{Tresse}, \& I.~{Vauglin}, 223--227

\bibitem[{{Allart} {et~al.}(2018){Allart}, {Bourrier}, {Lovis}, {Ehrenreich},
  {Spake}, {Wyttenbach}, {Pino}, {Pepe}, {Sing}, \& {Lecavelier des
  Etangs}}]{Allart_2018}
{Allart}, R., {Bourrier}, V., {Lovis}, C., {et~al.} 2018, Science, 362, 1384

\bibitem[{{Alonso-Floriano} {et~al.}(2019){Alonso-Floriano}, {Snellen},
  {Czesla}, {Bauer}, {Salz}, {Lamp{\'o}n}, {Lara}, {Nagel},
  {L{\'o}pez-Puertas}, {Nortmann}, {S{\'a}nchez-L{\'o}pez}, {Sanz-Forcada},
  {Caballero}, {Reiners}, {Ribas}, {Quirrenbach}, {Amado}, {Aceituno},
  {Anglada-Escud{\'e}}, {B{\'e}jar}, {Brinkm{\"o}ller}, {Hatzes}, {Henning},
  {Kaminski}, {K{\"u}rster}, {Labarga}, {Montes}, {Pall{\'e}}, {Schmitt}, \&
  {Zapatero Osorio}}]{Alonso_2019}
{Alonso-Floriano}, F.~J., {Snellen}, I.~A.~G., {Czesla}, S., {et~al.} 2019,
  \aap, 629, A110

\bibitem[{{Bailey} {et~al.}(2009){Bailey}, {Butler}, {Tinney}, {Jones},
  {O'Toole}, {Carter}, \& {Marcy}}]{Bailey_2009}
{Bailey}, J., {Butler}, R.~P., {Tinney}, C.~G., {et~al.} 2009, \apj, 690, 743

\bibitem[{{Bakos} {et~al.}(2010){Bakos}, {Torres}, {P{\'a}l}, {Hartman},
  {Kov{\'a}cs}, {Noyes}, {Latham}, {Sasselov}, {Sip{\H{o}}cz}, {Esquerdo},
  {Fischer}, {Johnson}, {Marcy}, {Butler}, {Isaacson}, {Howard}, {Vogt},
  {Kov{\'a}cs}, {Fernandez}, {Mo{\'o}r}, {Stefanik}, {L{\'a}z{\'a}r}, {Papp},
  \& {S{\'a}ri}}]{Bakos_2010}
{Bakos}, G.~{\'A}., {Torres}, G., {P{\'a}l}, A., {et~al.} 2010, \apj, 710, 1724

\bibitem[{{Ben-Jaffel}(2007)}]{Ben_Jaffel_2007}
{Ben-Jaffel}, L. 2007, \apjl, 671, L61

\bibitem[{{Ben-Jaffel} {et~al.}(2022){Ben-Jaffel}, {Ballester}, {Mu{\~n}oz},
  {Lavvas}, {Sing}, {Sanz-Forcada}, {Cohen}, {Kataria}, {Henry}, {Buchhave},
  {Mikal-Evans}, {Wakeford}, \& {L{\'o}pez-Morales}}]{Ben-Jaffel_2022}
{Ben-Jaffel}, L., {Ballester}, G.~E., {Mu{\~n}oz}, A.~G., {et~al.} 2022, Nature
  Astronomy, 6, 141

\bibitem[{Bennett {et~al.}(2023)Bennett, Redfield, Oklopčić, Carleo, Ninan,
  \& Endl}]{Bennett_2023}
Bennett, K.~A., Redfield, S., Oklopčić, A., {et~al.} 2023, The Astronomical
  Journal, 165, 264

\bibitem[{{Bonomo} {et~al.}(2017){Bonomo}, {Desidera}, {Benatti}, {Borsa},
  {Crespi}, {Damasso}, {Lanza}, {Sozzetti}, {Lodato}, {Marzari}, {Boccato},
  {Claudi}, {Cosentino}, {Covino}, {Gratton}, {Maggio}, {Micela}, {Molinari},
  {Pagano}, {Piotto}, {Poretti}, {Smareglia}, {Affer}, {Biazzo}, {Bignamini},
  {Esposito}, {Giacobbe}, {H{\'e}brard}, {Malavolta}, {Maldonado}, {Mancini},
  {Martinez Fiorenzano}, {Masiero}, {Nascimbeni}, {Pedani}, {Rainer}, \&
  {Scandariato}}]{Bonomo_2017}
{Bonomo}, A.~S., {Desidera}, S., {Benatti}, S., {et~al.} 2017, \aap, 602, A107

\bibitem[{{Bourrier} {et~al.}(2018){Bourrier}, {Lecavelier des Etangs},
  {Ehrenreich}, {Sanz-Forcada}, {Allart}, {Ballester}, {Buchhave}, {Cohen},
  {Deming}, {Evans}, {Garc{\'\i}a Mu{\~n}oz}, {Henry}, {Kataria}, {Lavvas},
  {Lewis}, {L{\'o}pez-Morales}, {Marley}, {Sing}, \&
  {Wakeford}}]{Bourrier_2018}
{Bourrier}, V., {Lecavelier des Etangs}, A., {Ehrenreich}, D., {et~al.} 2018,
  \aap, 620, A147

\bibitem[{Caldiroli {et~al.}(2021)Caldiroli, Haardt, Gallo, Spinelli, Malsky,
  \& Rauscher}]{Caldiroli_2021}
Caldiroli, A., Haardt, F., Gallo, E., {et~al.} 2021, A\&A, 655, A30

\bibitem[{{Caldiroli} {et~al.}(2022){Caldiroli}, {Haardt}, {Gallo}, {Spinelli},
  {Malsky}, \& {Rauscher}}]{Caldiroli_2022}
{Caldiroli}, A., {Haardt}, F., {Gallo}, E., {et~al.} 2022, \aap, 663, A122

\bibitem[{{Carolan} {et~al.}(2021){Carolan}, {Vidotto}, {Villarreal D'Angelo},
  \& {Hazra}}]{carolan21}
{Carolan}, S., {Vidotto}, A.~A., {Villarreal D'Angelo}, C., \& {Hazra}, G.
  2021, \mnras, 500, 3382

\bibitem[{{Chadney} {et~al.}(2015){Chadney}, {Galand}, {Unruh}, {Koskinen}, \&
  {Sanz-Forcada}}]{Chadney_2005}
{Chadney}, J.~M., {Galand}, M., {Unruh}, Y.~C., {Koskinen}, T.~T., \&
  {Sanz-Forcada}, J. 2015, \icarus, 250, 357

\bibitem[{{Cloutier} {et~al.}(2021){Cloutier}, {Charbonneau}, {Deming},
  {Bonfils}, \& {Astudillo-Defru}}]{Cloutier_2021}
{Cloutier}, R., {Charbonneau}, D., {Deming}, D., {Bonfils}, X., \&
  {Astudillo-Defru}, N. 2021, \aj, 162, 174

\bibitem[{{Craig} {et~al.}(1997){Craig}, {Abbott}, {Finley}, {Jessop},
  {Howell}, {Mathioudakis}, {Sommers}, {Vallerga}, \& {Malina}}]{craig97}
{Craig}, N., {Abbott}, M., {Finley}, D., {et~al.} 1997, \apjs, 113, 131

\bibitem[{{dos Santos} {et~al.}(2020){dos Santos}, {Ehrenreich}, {Bourrier},
  {Allart}, {King}, {Lendl}, {Lovis}, {Margheim}, {Mel{\'e}ndez}, {Seidel}, \&
  {Sousa}}]{dosSantos2020}
{dos Santos}, L.~A., {Ehrenreich}, D., {Bourrier}, V., {et~al.} 2020, \aap,
  640, A29

\bibitem[{Drake(1971)}]{Drake_1971}
Drake, G. W.~F. 1971, Phys. Rev. A, 3, 908

\bibitem[{{Duvvuri} {et~al.}(2021){Duvvuri}, {Sebastian Pineda},
  {Berta-Thompson}, {Brown}, {France}, {Kowalski}, {Redfield}, {Tilipman},
  {Vieytes}, {Wilson}, {Youngblood}, {Froning}, {Linsky}, {Parke Loyd},
  {Mauas}, {Miguel}, {Newton}, {Rugheimer}, \& {Christian
  Schneider}}]{Duvvuri_2021}
{Duvvuri}, G.~M., {Sebastian Pineda}, J., {Berta-Thompson}, Z.~K., {et~al.}
  2021, \apj, 913, 40

\bibitem[{{Ehrenreich} {et~al.}(2015){Ehrenreich}, {Bourrier}, {Wheatley},
  {Lecavelier des Etangs}, {H{\'e}brard}, {Udry}, {Bonfils}, {Delfosse},
  {D{\'e}sert}, {Sing}, \& {Vidal-Madjar}}]{Ehrenreich_2015}
{Ehrenreich}, D., {Bourrier}, V., {Wheatley}, P.~J., {et~al.} 2015, \nat, 522,
  459

\bibitem[{{Ellis} {et~al.}(2021){Ellis}, {Boyajian}, {von Braun}, {Ligi},
  {Mourard}, {Dragomir}, {Schaefer}, \& {Farrington}}]{Ellis_2021}
{Ellis}, T.~G., {Boyajian}, T., {von Braun}, K., {et~al.} 2021, \aj, 162, 118

\bibitem[{{Fortney} \& {Hubbard}(2004)}]{Fortney2004}
{Fortney}, J.~J. \& {Hubbard}, W.~B. 2004, \apj, 608, 1039

\bibitem[{{Fossati} {et~al.}(2022){Fossati}, {Guilluy}, {Shaikhislamov},
  {Carleo}, {Borsa}, {Bonomo}, {Giacobbe}, {Rainer}, {Cecchi-Pestellini},
  {Khodachenko}, {Efimov}, {Rumenskikh}, {Miroshnichenko}, {Berezutsky},
  {Nascimbeni}, {Brogi}, {Lanza}, {Mancini}, {Affer}, {Benatti}, {Biazzo},
  {Bignamini}, {Carosati}, {Claudi}, {Cosentino}, {Covino}, {Desidera},
  {Fiorenzano}, {Harutyunyan}, {Maggio}, {Malavolta}, {Maldonado}, {Micela},
  {Molinari}, {Pagano}, {Pedani}, {Piotto}, {Poretti}, {Scandariato},
  {Sozzetti}, \& {Stoev}}]{fossati22}
{Fossati}, L., {Guilluy}, G., {Shaikhislamov}, I.~F., {et~al.} 2022, \aap, 658,
  A136

\bibitem[{{Fossati} {et~al.}(2023){Fossati}, {Pillitteri}, {Shaikhislamov},
  {Bonfanti}, {Borsa}, {Carleo}, {Guilluy}, \& {Rumenskikh}}]{Fossati_2023}
{Fossati}, L., {Pillitteri}, I., {Shaikhislamov}, I.~F., {et~al.} 2023, \aap,
  673, A37

\bibitem[{{France} {et~al.}(2022){France}, {Fleming}, {Youngblood}, {Mason},
  {Drake}, {Amerstorfer}, {Barstow}, {Bourrier}, {Champey}, {Fossati},
  {Froning}, {Green}, {Gris{\'e}}, {Gronoff}, {Hellickson}, {Jin}, {Koskinen},
  {Kowalski}, {Kruczek}, {Linsky}, {Lipscy}, {McEntaffer}, {McKenzie}, {Miles},
  {Patton}, {Savage}, {Siegmund}, {Spittler}, {Unruh}, \& {Volz}}]{France_2022}
{France}, K., {Fleming}, B., {Youngblood}, A., {et~al.} 2022, Journal of
  Astronomical Telescopes, Instruments, and Systems, 8, 014006

\bibitem[{{France} {et~al.}(2016){France}, {Loyd}, {Youngblood}, {Brown},
  {Schneider}, {Hawley}, {Froning}, {Linsky}, {Roberge}, {Buccino},
  {Davenport}, {Fontenla}, {Kaltenegger}, {Kowalski}, {Mauas}, {Miguel},
  {Redfield}, {Rugheimer}, {Tian}, {Vieytes}, {Walkowicz}, \&
  {Weisenburger}}]{France_2016}
{France}, K., {Loyd}, R.~O.~P., {Youngblood}, A., {et~al.} 2016, \apj, 820, 89

\bibitem[{{Fulton} {et~al.}(2017){Fulton}, {Petigura}, {Howard}, {Isaacson},
  {Marcy}, {Cargile}, {Hebb}, {Weiss}, {Johnson}, {Morton}, {Sinukoff},
  {Crossfield}, \& {Hirsch}}]{fulton17}
{Fulton}, B.~J., {Petigura}, E.~A., {Howard}, A.~W., {et~al.} 2017, \aj, 154,
  109

\bibitem[{{Guilluy} {et~al.}(2020){Guilluy}, {Andretta}, {Borsa}, {Giacobbe},
  {Sozzetti}, {Covino}, {Bourrier}, {Fossati}, {Bonomo}, {Esposito},
  {Giampapa}, {Harutyunyan}, {Rainer}, {Brogi}, {Bruno}, {Claudi}, {Frustagli},
  {Lanza}, {Mancini}, {Pino}, {Poretti}, {Scandariato}, {Affer}, {Baffa},
  {Baruffolo}, {Benatti}, {Biazzo}, {Bignamini}, {Boschin}, {Carleo},
  {Cecconi}, {Cosentino}, {Damasso}, {Desidera}, {Falcini}, {Martinez
  Fiorenzano}, {Ghedina}, {Gonz{\'a}lez-{\'A}lvarez}, {Guerra}, {Hernandez},
  {Leto}, {Maggio}, {Malavolta}, {Maldonado}, {Micela}, {Molinari},
  {Nascimbeni}, {Pagano}, {Pedani}, {Piotto}, \& {Reiners}}]{Guilluy_2020}
{Guilluy}, G., {Andretta}, V., {Borsa}, F., {et~al.} 2020, \aap, 639, A49

\bibitem[{{Guilluy} {et~al.}(2023){Guilluy}, {Bourrier}, {Jaziri}, {Dethier},
  {Mounzer}, {Giacobbe}, {Attia}, {Allart}, {Bonomo}, {Dos Santos}, {Rainer},
  \& {Sozzetti}}]{Guilluy_2023}
{Guilluy}, G., {Bourrier}, V., {Jaziri}, Y., {et~al.} 2023, arXiv e-prints,
  arXiv:2307.00967

\bibitem[{{Husser} {et~al.}(2013){Husser}, {Wende-von Berg}, {Dreizler},
  {Homeier}, {Reiners}, {Barman}, \& {Hauschildt}}]{Husser_2013}
{Husser}, T.~O., {Wende-von Berg}, S., {Dreizler}, S., {et~al.} 2013, \aap,
  553, A6

\bibitem[{{Kasper} {et~al.}(2020){Kasper}, {Bean}, {Oklop{\v{c}}i{\'c}},
  {Malsky}, {Kempton}, {D{\'e}sert}, {Rogers}, \& {Mansfield}}]{Kasper2020}
{Kasper}, D., {Bean}, J.~L., {Oklop{\v{c}}i{\'c}}, A., {et~al.} 2020, \aj, 160,
  258

\bibitem[{{King} {et~al.}(2018){King}, {Wheatley}, {Salz}, {Bourrier},
  {Czesla}, {Ehrenreich}, {Kirk}, {Lecavelier des Etangs}, {Louden}, {Schmitt},
  \& {Schneider}}]{King_2018}
{King}, G.~W., {Wheatley}, P.~J., {Salz}, M., {et~al.} 2018, \mnras, 478, 1193

\bibitem[{{Koskinen} {et~al.}(2014){Koskinen}, {Lavvas}, {Harris}, \&
  {Yelle}}]{Koskinen2014}
{Koskinen}, T.~T., {Lavvas}, P., {Harris}, M.~J., \& {Yelle}, R.~V. 2014,
  Philosophical Transactions of the Royal Society of London Series A, 372,
  20130089

\bibitem[{{Kreidberg} \& {Oklop{\v{c}}i{\'c}}(2018)}]{Kreidberg2018}
{Kreidberg}, L. \& {Oklop{\v{c}}i{\'c}}, A. 2018, Research Notes of the
  American Astronomical Society, 2, 44

\bibitem[{{Kubyshkina} {et~al.}(2018){Kubyshkina}, {Fossati}, {Erkaev},
  {Cubillos}, {Johnstone}, {Kislyakova}, {Lammer}, {Lendl}, \&
  {Odert}}]{Kubyshkina_2018}
{Kubyshkina}, D., {Fossati}, L., {Erkaev}, N.~V., {et~al.} 2018, \apjl, 866,
  L18

\bibitem[{Kulow {et~al.}(2014)Kulow, France, Linsky, \& Loyd}]{Kulow_2014}
Kulow, J.~R., France, K., Linsky, J., \& Loyd, R. O.~P. 2014, The Astrophysical
  Journal, 786, 132

\bibitem[{Lamp\'on {et~al.}(2020)Lamp\'on, L\'opez-Puertas, Lara,
  S\'anchez-L\'opez, Salz, Czesla, Sanz-Forcada, Molaverdikhani,
  Alonso-Floriano, Nortmann, Caballero, Bauer, Pall\'e, Montes, Quirrenbach,
  Nagel, Ribas, Reiners, \& Amado}]{Lampon2020}
Lamp\'on, M., L\'opez-Puertas, M., Lara, L.~M., {et~al.} 2020, A\&A, 636, A13

\bibitem[{{Lecavelier Des Etangs} {et~al.}(2010){Lecavelier Des Etangs},
  {Ehrenreich}, {Vidal-Madjar}, {Ballester}, {D{\'e}sert}, {Ferlet},
  {H{\'e}brard}, {Sing}, {Tchakoumegni}, \& {Udry}}]{Lecavelier_2010}
{Lecavelier Des Etangs}, A., {Ehrenreich}, D., {Vidal-Madjar}, A., {et~al.}
  2010, \aap, 514, A72

\bibitem[{{Linsky} {et~al.}(2014){Linsky}, {Fontenla}, \&
  {France}}]{Linsky_2014}
{Linsky}, J.~L., {Fontenla}, J., \& {France}, K. 2014, \apj, 780, 61

\bibitem[{{Linssen} {et~al.}(2022){Linssen}, {Oklop{\v{c}}i{\'c}}, \&
  {MacLeod}}]{linssen22}
{Linssen}, D.~C., {Oklop{\v{c}}i{\'c}}, A., \& {MacLeod}, M. 2022, \aap, 667,
  A54

\bibitem[{{Lopez} \& {Fortney}(2014)}]{Lopez2014}
{Lopez}, E.~D. \& {Fortney}, J.~J. 2014, \apj, 792, 1

\bibitem[{{Loyd} {et~al.}(2016){Loyd}, {France}, {Youngblood}, {Schneider},
  {Brown}, {Hu}, {Linsky}, {Froning}, {Redfield}, {Rugheimer}, \&
  {Tian}}]{Loyd_2016}
{Loyd}, R.~O.~P., {France}, K., {Youngblood}, A., {et~al.} 2016, \apj, 824, 102

\bibitem[{{MacLeod} \& {Oklop{\v{c}}i{\'c}}(2022)}]{macleod}
{MacLeod}, M. \& {Oklop{\v{c}}i{\'c}}, A. 2022, \apj, 926, 226

\bibitem[{{Malsky} {et~al.}(2023){Malsky}, {Rogers}, {Kempton}, \&
  {Marounina}}]{Malsky_2023}
{Malsky}, I., {Rogers}, L., {Kempton}, E. M.~R., \& {Marounina}, N. 2023,
  Nature Astronomy, 7, 57

\bibitem[{{Mansfield} {et~al.}(2018){Mansfield}, {Bean}, {Oklop{\v{c}}i{\'c}},
  {Kreidberg}, {D{\'e}sert}, {Kempton}, {Line}, {Fortney}, {Henry}, {Mallonn},
  {Stevenson}, {Dragomir}, {Allart}, \& {Bourrier}}]{Mansfield_2018}
{Mansfield}, M., {Bean}, J.~L., {Oklop{\v{c}}i{\'c}}, A., {et~al.} 2018, \apjl,
  868, L34

\bibitem[{{Ninan} {et~al.}(2020){Ninan}, {Stefansson}, {Mahadevan}, {Bender},
  {Robertson}, {Ramsey}, {Terrien}, {Wright}, {Diddams}, {Kanodia}, {Cochran},
  {Endl}, {Ford}, {Fredrick}, {Halverson}, {Hearty}, {Jennings}, {Kaplan},
  {Lubar}, {Metcalf}, {Monson}, {Nitroy}, {Roy}, \& {Schwab}}]{Ninan2020}
{Ninan}, J.~P., {Stefansson}, G., {Mahadevan}, S., {et~al.} 2020, \apj, 894, 97

\bibitem[{Norcross(1971)}]{Norcross1971}
Norcross, D.~W. 1971, Journal of Physics B: Atomic and Molecular Physics, 4,
  652

\bibitem[{{Nortmann} {et~al.}(2018){Nortmann}, {Pall{\'e}}, {Salz},
  {Sanz-Forcada}, {Nagel}, {Alonso-Floriano}, {Czesla}, {Yan}, {Chen},
  {Snellen}, {Zechmeister}, {Schmitt}, {L{\'o}pez-Puertas}, {Casasayas-Barris},
  {Bauer}, {Amado}, {Caballero}, {Dreizler}, {Henning}, {Lamp{\'o}n}, {Montes},
  {Molaverdikhani}, {Quirrenbach}, {Reiners}, {Ribas}, {S{\'a}nchez-L{\'o}pez},
  {Schneider}, \& {Zapatero Osorio}}]{Nortmann_2018}
{Nortmann}, L., {Pall{\'e}}, E., {Salz}, M., {et~al.} 2018, Science, 362, 1388

\bibitem[{{Oklop{\v{c}}i{\'c}} \& {Hirata}(2018)}]{Oklopcic_2018}
{Oklop{\v{c}}i{\'c}}, A. \& {Hirata}, C.~M. 2018, \apjl, 855, L11

\bibitem[{{Oklop{\v{c}}i{\'c}} {et~al.}(2020){Oklop{\v{c}}i{\'c}}, {Silva},
  {Montero-Camacho}, \& {Hirata}}]{Oklopcic2020}
{Oklop{\v{c}}i{\'c}}, A., {Silva}, M., {Montero-Camacho}, P., \& {Hirata},
  C.~M. 2020, \apj, 890, 88

\bibitem[{Oklopčić(2019)}]{Oklopcic_2019}
Oklopčić, A. 2019, The Astrophysical Journal, 881, 133

\bibitem[{{Orell-Miquel} {et~al.}(2022){Orell-Miquel}, {Murgas}, {Pall{\'e}},
  {Lamp{\'o}n}, {L{\'o}pez-Puertas}, {Sanz-Forcada}, {Nagel}, {Kaminski},
  {Casasayas-Barris}, {Nortmann}, {Luque}, {Molaverdikhani}, {Sedaghati},
  {Caballero}, {Amado}, {Bergond}, {Czesla}, {Hatzes}, {Henning},
  {Khalafinejad}, {Montes}, {Morello}, {Quirrenbach}, {Reiners}, {Ribas},
  {S{\'a}nchez-L{\'o}pez}, {Schweitzer}, {Stangret}, {Yan}, \& {Zapatero
  Osorio}}]{Orell-Miquel2022}
{Orell-Miquel}, J., {Murgas}, F., {Pall{\'e}}, E., {et~al.} 2022, \aap, 659,
  A55

\bibitem[{{Owen} {et~al.}(2023){Owen}, {Murray-Clay}, {Schreyer},
  {Schlichting}, {Ardila}, {Gupta}, {Loyd}, {Shkolnik}, {Sing}, \&
  {Swain}}]{Owen_2023}
{Owen}, J.~E., {Murray-Clay}, R.~A., {Schreyer}, E., {et~al.} 2023, \mnras,
  518, 4357

\bibitem[{{Owen} \& {Wu}(2013)}]{owenwu13}
{Owen}, J.~E. \& {Wu}, Y. 2013, \apj, 775, 105

\bibitem[{{Owen} \& {Wu}(2017)}]{owenwu17}
{Owen}, J.~E. \& {Wu}, Y. 2017, \apj, 847, 29

\bibitem[{{Palle} {et~al.}(2020){Palle}, {Nortmann}, {Casasayas-Barris},
  {Lamp{\'o}n}, {L{\'o}pez-Puertas}, {Caballero}, {Sanz-Forcada}, {Lara},
  {Nagel}, {Yan}, {Alonso-Floriano}, {Amado}, {Chen}, {Cifuentes},
  {Cort{\'e}s-Contreras}, {Czesla}, {Molaverdikhani}, {Montes}, {Passegger},
  {Quirrenbach}, {Reiners}, {Ribas}, {S{\'a}nchez-L{\'o}pez}, {Schweitzer},
  {Stangret}, {Zapatero Osorio}, \& {Zechmeister}}]{Palle2020}
{Palle}, E., {Nortmann}, L., {Casasayas-Barris}, N., {et~al.} 2020, \aap, 638,
  A61

\bibitem[{{Paragas} {et~al.}(2021){Paragas}, {Vissapragada}, {Knutson},
  {Oklop{\v{c}}i{\'c}}, {Chachan}, {Greklek-McKeon}, {Dai}, {Tinyanont}, \&
  {Vasisht}}]{Paragas2021}
{Paragas}, K., {Vissapragada}, S., {Knutson}, H.~A., {et~al.} 2021, \apjl, 909,
  L10

\bibitem[{{Paredes} {et~al.}(2021){Paredes}, {Henry}, {Quinn}, {Gies},
  {Hinojosa-Go{\~n}i}, {James}, {Jao}, \& {White}}]{Paredes_2021}
{Paredes}, L.~A., {Henry}, T.~J., {Quinn}, S.~N., {et~al.} 2021, \aj, 162, 176

\bibitem[{{Parker}(1958)}]{Parker_1958}
{Parker}, E.~N. 1958, \apj, 128, 664

\bibitem[{{Pepe} {et~al.}(2011){Pepe}, {Lovis}, {S{\'e}gransan}, {Benz},
  {Bouchy}, {Dumusque}, {Mayor}, {Queloz}, {Santos}, \& {Udry}}]{Pepe_2011}
{Pepe}, F., {Lovis}, C., {S{\'e}gransan}, D., {et~al.} 2011, \aap, 534, A58

\bibitem[{{Pillitteri} {et~al.}(2022){Pillitteri}, {Micela, G.}, {Maggio, A.},
  {Sciortino, S.}, \& {Lopez-Santiago, J.}}]{Pillitteri_2022}
{Pillitteri}, {Micela, G.}, {Maggio, A.}, {Sciortino, S.}, \& {Lopez-Santiago,
  J.} 2022, A\&A, 660, A75

\bibitem[{{Poppenhaeger}(2022)}]{Poppenhaeger_2022}
{Poppenhaeger}, K. 2022, \mnras, 512, 1751

\bibitem[{{Salz} {et~al.}(2015){Salz}, {Banerjee}, {Mignone}, {Schneider},
  {Czesla}, \& {Schmitt}}]{Salz2015}
{Salz}, M., {Banerjee}, R., {Mignone}, A., {et~al.} 2015, \aap, 576, A21

\bibitem[{{Salz} {et~al.}(2018){Salz}, {Czesla}, {Schneider}, {Nagel},
  {Schmitt}, {Nortmann}, {Alonso-Floriano}, {L{\'o}pez-Puertas}, {Lamp{\'o}n},
  {Bauer}, {Snellen}, {Pall{\'e}}, {Caballero}, {Yan}, {Chen}, {Sanz-Forcada},
  {Amado}, {Quirrenbach}, {Ribas}, {Reiners}, {B{\'e}jar}, {Casasayas-Barris},
  {Cort{\'e}s-Contreras}, {Dreizler}, {Guenther}, {Henning}, {Jeffers},
  {Kaminski}, {K{\"u}rster}, {Lafarga}, {Lara}, {Molaverdikhani}, {Montes},
  {Morales}, {S{\'a}nchez-L{\'o}pez}, {Seifert}, {Zapatero Osorio}, \&
  {Zechmeister}}]{Salz_2018}
{Salz}, M., {Czesla}, S., {Schneider}, P.~C., {et~al.} 2018, \aap, 620, A97

\bibitem[{{Salz} {et~al.}(2016{\natexlab{a}}){Salz}, {Czesla}, {Schneider}, \&
  {Schmitt}}]{Salz2016a}
{Salz}, M., {Czesla}, S., {Schneider}, P.~C., \& {Schmitt}, J.~H.~M.~M.
  2016{\natexlab{a}}, \aap, 586, A75

\bibitem[{{Salz} {et~al.}(2016{\natexlab{b}}){Salz}, {Schneider}, {Czesla}, \&
  {Schmitt}}]{Salz2016b}
{Salz}, M., {Schneider}, P., {Czesla}, S., \& {Schmitt}, J.~H.~M.~M.
  2016{\natexlab{b}}, \aap, 585, L2

\bibitem[{{S{\'a}nchez-L{\'o}pez} {et~al.}(2022){S{\'a}nchez-L{\'o}pez}, {Lin},
  {Snellen}, {Casasayas-Barris}, {Garc{\'\i}a Mu{\~n}oz}, {Lamp{\'o}n}, \&
  {L{\'o}pez-Puertas}}]{Sanchez2022}
{S{\'a}nchez-L{\'o}pez}, A., {Lin}, L., {Snellen}, I.~A.~G., {et~al.} 2022,
  \aap, 666, L1

\bibitem[{{Schreyer} {et~al.}(2023){Schreyer}, {Owen}, {Spake}, {Bahroloom}, \&
  {Di Giampasquale}}]{Schreyer_2023}
{Schreyer}, E., {Owen}, J.~E., {Spake}, J.~J., {Bahroloom}, Z., \& {Di
  Giampasquale}, S. 2023, arXiv e-prints, arXiv:2302.10947

\bibitem[{{Seager} \& {Sasselov}(2000)}]{Seager_2000}
{Seager}, S. \& {Sasselov}, D.~D. 2000, \apj, 537, 916

\bibitem[{{Spake} {et~al.}(2018){Spake}, {Sing}, {Evans}, {Oklop{\v{c}}i{\'c}},
  {}, {Bourrier}, {Kreidberg}, {Rackham}, {Irwin}, {Ehrenreich}, {Wyttenbach},
  {Wakeford}, {Zhou}, {Chubb}, {Nikolov}, {Goyal}, {Henry}, {Williamson},
  {Blumenthal}, {Anderson}, {Hellier}, {Charbonneau}, {Udry}, \&
  {Madhusudhan}}]{Spake_2018}
{Spake}, J.~J., {Sing}, D.~K., {Evans}, T.~M., {et~al.} 2018, \nat, 557, 68

\bibitem[{{Stassun} {et~al.}(2019){Stassun}, {Oelkers}, {Paegert}, {Torres},
  {Pepper}, {De Lee}, {Collins}, {Latham}, {Muirhead}, {Chittidi},
  {Rojas-Ayala}, {Fleming}, {Rose}, {Tenenbaum}, {Ting}, {Kane}, {Barclay},
  {Bean}, {Brassuer}, {Charbonneau}, {Ge}, {Lissauer}, {Mann}, {McLean},
  {Mullally}, {Narita}, {Plavchan}, {Ricker}, {Sasselov}, {Seager}, {Sharma},
  {Shiao}, {Sozzetti}, {Stello}, {Vanderspek}, {Wallace}, \&
  {Winn}}]{Stassun_2019}
{Stassun}, K.~G., {Oelkers}, R.~J., {Paegert}, M., {et~al.} 2019, \aj, 158, 138

\bibitem[{Tian {et~al.}(2005)Tian, Toon, Pavlov, \& De~Sterck}]{Tian2005}
Tian, F., Toon, O., Pavlov, A., \& De~Sterck, H. 2005, \apj, 621, 1049

\bibitem[{{Tuomi} {et~al.}(2013){Tuomi}, {Anglada-Escud{\'e}}, {Gerlach},
  {Jones}, {Reiners}, {Rivera}, {Vogt}, \& {Butler}}]{Tuomi_2013}
{Tuomi}, M., {Anglada-Escud{\'e}}, G., {Gerlach}, E., {et~al.} 2013, \aap, 549,
  A48

\bibitem[{{Vidal-Madjar} {et~al.}(2003){Vidal-Madjar}, {Lecavelier des Etangs},
  {D{\'e}sert}, {Ballester}, {Ferlet}, {H{\'e}brard}, \& {Mayor}}]{Vidal_2003}
{Vidal-Madjar}, A., {Lecavelier des Etangs}, A., {D{\'e}sert}, J.~M., {et~al.}
  2003, \nat, 422, 143

\bibitem[{{Vidotto} \& {Cleary}(2020)}]{Vidotto_2020}
{Vidotto}, A.~A. \& {Cleary}, A. 2020, \mnras, 494, 2417

\bibitem[{{Vissapragada} {et~al.}(2022){Vissapragada}, {Knutson}, {dos Santos},
  {Wang}, \& {Dai}}]{vissa22}
{Vissapragada}, S., {Knutson}, H.~A., {dos Santos}, L.~A., {Wang}, L., \&
  {Dai}, F. 2022, \apj, 927, 96

\bibitem[{{Vissapragada} {et~al.}(2021){Vissapragada}, {Stef{\'a}nsson},
  {Greklek-McKeon}, {Oklop{\v{c}}i{\'c}}, {Knutson}, {Ninan}, {Mahadevan},
  {Ca{\~n}as}, {Chachan}, {Cochran}, {Collins}, {Dai}, {David}, {Halverson},
  {Hawley}, {Hebb}, {Kanodia}, {Kowalski}, {Livingston}, {Maney}, {Metcalf},
  {Morley}, {Ramsey}, {Robertson}, {Roy}, {Spake}, {Schwab}, {Terrien},
  {Tinyanont}, {Vasisht}, \& {Wisniewski}}]{vissa21}
{Vissapragada}, S., {Stef{\'a}nsson}, G., {Greklek-McKeon}, M., {et~al.} 2021,
  \aj, 162, 222

\bibitem[{{Yan} {et~al.}(2022){Yan}, {Seon}, {Guo}, {Chen}, \& {Li}}]{Yan_2022}
{Yan}, D., {Seon}, K.-i., {Guo}, J., {Chen}, G., \& {Li}, L. 2022, \apj, 936,
  177

\bibitem[{Yelle(2004)}]{Yelle2004}
Yelle, R.~V. 2004, \icarus, 170, 167

\bibitem[{{Youngblood} {et~al.}(2016){Youngblood}, {France}, {Loyd}, {Linsky},
  {Redfield}, {Schneider}, {Wood}, {Brown}, {Froning}, {Miguel}, {Rugheimer},
  \& {Walkowicz}}]{Youngblood_2016}
{Youngblood}, A., {France}, K., {Loyd}, R.~O.~P., {et~al.} 2016, \apj, 824, 101

\bibitem[{{Zhang} {et~al.}(2022){Zhang}, {Knutson}, {Wang}, {Dai}, \&
  {Barrag{\'a}n}}]{Zhang2022}
{Zhang}, M., {Knutson}, H.~A., {Wang}, L., {Dai}, F., \& {Barrag{\'a}n}, O.
  2022, \aj, 163, 67

\bibitem[{{Zhang} {et~al.}(2021){Zhang}, {Knutson}, {Wang}, {Dai}, {Oklopcic},
  \& {Hu}}]{Zhang2021}
{Zhang}, M., {Knutson}, H.~A., {Wang}, L., {et~al.} 2021, \aj, 161, 181

\bibitem[{{Zhang} {et~al.}(2023){Zhang}, {Morley}, {Gully-Santiago}, {MacLeod},
  {Oklop{\v{c}}i{\'c}}, {Luna}, {Tran}, {Ninan}, {Mahadevan}, {Krolikowski},
  {Cochran}, {Bowler}, {Endl}, {Stef{\'a}nsson}, {Tofflemire}, {Vanderburg}, \&
  {Zeimann}}]{Zhang_2023}
{Zhang}, Z., {Morley}, C.~V., {Gully-Santiago}, M., {et~al.} 2023, Science
  Advances, 9, eadf8736

\end{thebibliography}

\begin{appendix}
\section{Reaction Rates}
\label{appendix:rates}
\begin{figure}[!htb]
    \centering
    \onecolumn
        \includegraphics[width = .7\textwidth]{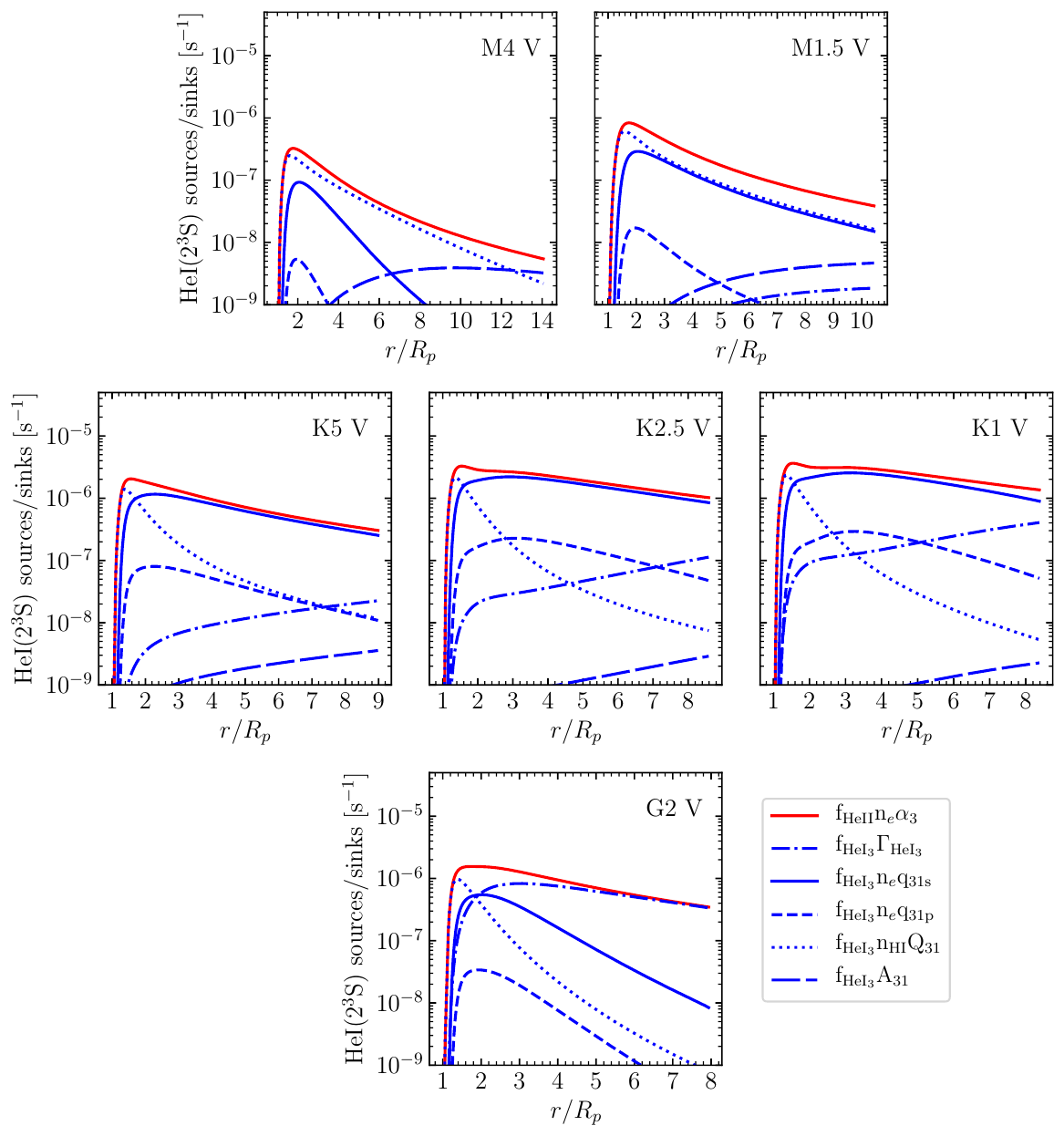}
        \caption{Reaction rates for the population/depopulation of the \HeITR state for the simulations of HAT-P-11b shown in Figure \ref{fig:HATP11_types} (both the $q_{13}$ and advection rates are lower than $10^{-9}$ s$^{-1}$).}
        \label{fig:a3}
\end{figure}
\twocolumn
\begin{figure*}
    \centering
    \includegraphics[width = .7\textwidth]{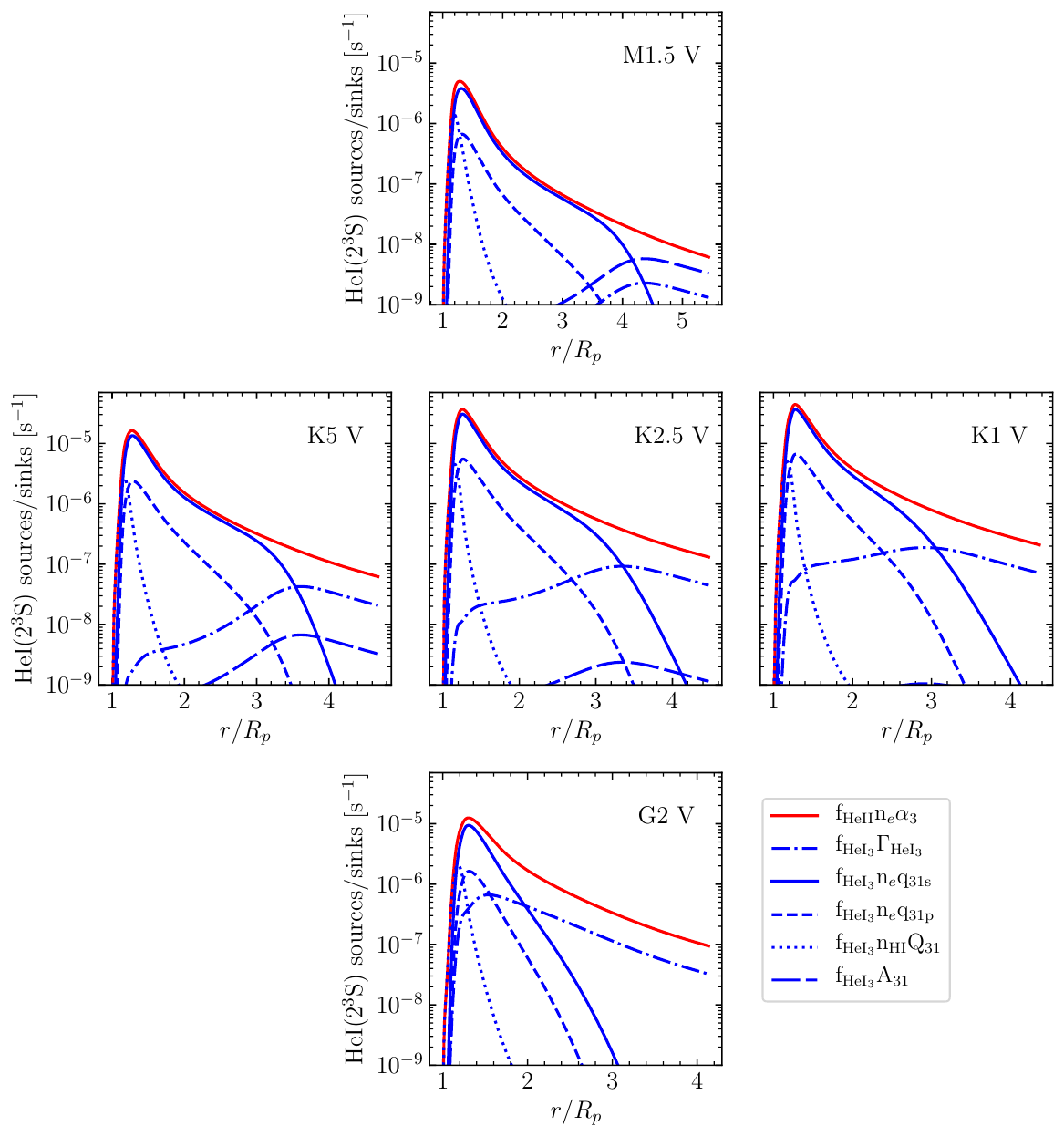}
    \caption{Same as Figure \ref{fig:a3}, but for the simulations of HD~189733~b shown in Figure \ref{fig:HD189733_types}. }
    \label{fig:a4}
\end{figure*}
\end{appendix}

\end{document}